\documentstyle[12pt,epsfig]{article}
\newskip\humongous \humongous=0pt plus 1000pt minus 1000pt

\newif\ifdtup

\def\ie{\hbox{\rm i.e.}{}}

\def\Im{\mathop{\rm Im}}
\def\Re{\mathop{\rm Re}}

\def\pr#1{#1^\prime}

\def\beq{\begin{equation}}
\def\eeq{\end{equation}}
\def\eq{\beq\eeq}
\def\beqn{\begin{eqnarray}}
\def\eeqn{\end{eqnarray}}
\def\dispfrac#1#2{{\displaystyle #1 \over \displaystyle #2}}
\relax

\def\dotx{\dotx{\dot\overline{x}}}

\relax

\catcode`\@=11

\@addtoreset{equation}{section}
\def\theequation{\arabic{section}.\arabic{equation}}

\def\@normalsize{\@setsize\normalsize{15pt}\xiipt\@xiipt
\abovedisplayskip 14pt plus3pt minus3pt%
\belowdisplayskip \abovedisplayskip
\abovedisplayshortskip \z@ plus3pt%
\belowdisplayshortskip 7pt plus3.5pt minus0pt}

\def\small{\@setsize\small{13.6pt}\xipt\@xipt
\abovedisplayskip 13pt plus3pt minus3pt%
\belowdisplayskip \abovedisplayskip
\abovedisplayshortskip \z@ plus3pt%
\belowdisplayshortskip 7pt plus3.5pt minus0pt
\def\@listi{\parsep 4.5pt plus 2pt minus 1pt
     \itemsep \parsep
     \topsep 9pt plus 3pt minus 3pt}}

\@twosidetrue

\relax

\catcode`@=12

\catcode`\@=11

\def\section{\@startsection{section}{1}{\z@}{3.5ex plus 1ex minus
   .2ex}{2.3ex plus .2ex}{\large\bf}}

\def\thesection{\arabic{section}}

\def\appendix{\setcounter{section}{0}
 \def\thesection{APPENDIX \Alph{section}:}
 \def\theequation{\Alph{section}.\arabic{equation}}}

\def\ps@headings{\def\@oddfoot{}\def\@evenfoot{}
\def\@oddhead{\hbox{}\hfill
 \makebox[.5\textwidth]{\raggedright\ignorespaces --\thepage{}--
 \hfill {}}}  
\def\@evenhead{\@oddhead}
\def\subsectionmark##1{\markboth{##1}{}}
}

\ps@headings

\catcode`\@=12

\def\figcap{\section*{Figure Captions\markboth
 {FIGURECAPTIONS}{FIGURECAPTIONS}}\list
 {Fig. \arabic{enumi}:\hfill}{\settowidth\labelwidth{Fig. 999:}
 \leftmargin\labelwidth
 \advance\leftmargin\labelsep\usecounter{enumi}}}
 \relax
\def\tablecap{\section*{Table Captions\markboth
 {TABLECAPTIONS}{TABLECAPTIONS}}\list
 {Table \arabic{enumi}:\hfill}{\settowidth\labelwidth{Table 999:}
 \leftmargin\labelwidth
 \advance\leftmargin\labelsep\usecounter{enumi}}}
 \relax
\def\reflist{\section*{References\markboth
 {REFLIST}{REFLIST}}\list
 {[\arabic{enumi}]\hfill}{\settowidth\labelwidth{[999]}
 \leftmargin\labelwidth
 \advance\leftmargin\labelsep\usecounter{enumi}}}
 \relax

\catcode`\@=11

\def\ps@headings{\def\@oddfoot{}\def\@evenfoot{}
\def\@oddhead{\hbox{}\hfill
 \makebox[.5\textwidth]{\raggedright\ignorespaces --\thepage{}--
 \hfill {}}}    
\def\@evenhead{\@oddhead}
\def\subsectionmark##1{\markboth{##1}{}}
}

\ps@headings

\relax

\relax
\def\pl#1#2#3{{\it Phys. Lett. }{\bf #1}(19#2)#3}

\def\prl#1#2#3{{\it Phys. Rev. Lett. }{\bf #1}(19#2)#3}

\def\prep#1#2#3{{\it Phys. Rep. }{\bf #1}(19#2)#3}
\def\pr#1#2#3{{\it Phys. Rev. }{\bf #1}(19#2)#3}
\def\np#1#2#3{{\it Nucl. Phys. }{\bf #1}(19#2)#3}

\relax

\catcode `@ 11
\def\biblabel#1{\if@filesw\immediate
\write\@auxout{\string\bibcite{#1}{\the\value{\@listctr }}}\fi}
\catcode `@ 12

\begin{document}
\renewcommand\floatpagefraction{0.0}
\setcounter{topnumber}{5}
\renewcommand\topfraction{1.0}
\setcounter{bottomnumber}{5}
\renewcommand\bottomfraction{1.0}
\renewcommand\textfraction{0.0}
\setcounter{totalnumber}{5}        
\setcounter{dbltopnumber}{2}       
\newcommand\ssmallfig{3cm}
\newcommand\smallfig{4cm}
\newcommand\mediumfig{5cm}
\newcommand\bigfig{6cm}
\newcommand\bbigfig{9cm}
\newcommand\as{\alpha_S}
\newcommand\aem{\alpha_{\rm em}}
\newcommand\refq[1]{$^{[#1]}$}
\newcommand\avr[1]{\left\langle #1 \right\rangle}
\newcommand\lambdamsb{\ifmmode
\Lambda_5^{\rm \scriptscriptstyle \overline{MS}}
\else
$\Lambda_5^{\rm \scriptscriptstyle \overline{MS}}$
\fi}
\newcommand\nf{\ifmmode n_{\rm f} \else $n_{\rm f}$ \fi}
\newcommand\MSB{\ifmmode \overline{\rm MS} \else $\overline{\rm MS}$ \fi}
\newcommand\qqb{{q\overline{q}}}
\newcommand\asb{\as^{(b)}}
\newcommand\qb{\overline{q}}
\newcommand\sigqq{\sigma_{q\overline{q}}}
\newcommand\siggg{\sigma_{gg}}
\newcommand\ep{\epsilon}
\def\Im{\mathop{\rm Im}}
\def\Re{\mathop{\rm Re}}
\def\bzero{b_0}
\def\bone{b_1}
\def\half{\frac{1}{2}}
\def\GE{\gamma_E}
\def\asp{{\alpha_S}\over{\pi}}
\def \eq {e_{\scriptscriptstyle Q}}
\def \muf {\mu_{\scriptscriptstyle F}}
\def \mur {\mu_{\scriptscriptstyle R}}
\def \muo {\mbox{$\mu_0$}}
\def \pt  {\mbox{$p_{\rm T}$}}
\def \ptg {\mbox{$p_{\rm T}^{q\overline{q}}$}}
\def \xf  {\mbox{$x_{\rm F}$}}
\def \dphi{\mbox{$\Delta\phi$}}
\def \dy  {\mbox{$\Delta y$}}
\def \pim {\mbox{$\pi^-$}}
\def \epem {\mbox{$e^+e^-$}}
\def \mc   {\mbox{$m_c$}}
\def \mb   {\mbox{$m_b$}}
\def \mqq   {\mbox{$M_{q\overline{q}}$}}
\def \tot   {{\rm tot}}
\newcommand\qq{{\scriptscriptstyle Q\overline{Q}}}
\newcommand\cm{{\scriptscriptstyle CM}}
\def\beeq#1{\begin{eqnarray}\label{#1}}
\def\eeeq{\end{eqnarray}}
\newcommand{\ccaption}[2]{
  \begin{center}
    \parbox{0.85\textwidth}{
      \caption[#1]{\small\it {#2}}}
  \end{center}    }
%
\begin{titlepage}
\nopagebreak
\vspace*{-1in}
{\leftskip 11cm
\normalsize
\noindent
\newline
CERN-TH/96-86 \\
hep-ph/9604351

}
\begin{center}
{\large \bf \sc The Resummation of soft gluons\\ in Hadronic Collisions}
\vskip 1cm
{\bf Stefano CATANI\footnote{Research supported in part by the EEC programme
``Human Capital and Mobility'', Network ``Physics at High Energy Colliders'',
contract CHRX-CT93-0357(DG 12 COMA).}}
\\
\vskip .1cm
{INFN, Sezione di Firenze and Univ. di Firenze, Florence, Italy} \\
\vskip .5cm
{\bf Michelangelo L. MANGANO\footnote{On leave of absence
from INFN, Pisa, Italy.}} and
{\bf Paolo NASON\footnote{On leave of absence from INFN, Milan, Italy.}}
\\
\vskip 0.1cm
{CERN, TH Division, Geneva, Switzerland} \\
\vskip .5cm
{\bf Luca TRENTADUE$^1$}
\\
\vskip .1cm
{Univ. di Parma and INFN, Gruppo Collegato di Parma, Parma, Italy}
\end{center}
\vskip 0.5cm
\nopagebreak
\begin{abstract}
{\small
We compute the effects of soft gluon resummation for
the production of high mass systems in hadronic collisions.
We carefully analyse the growth of the perturbative expansion coefficients
of the resummation formula. We propose an expression
consistent with the known leading and next-to-leading resummation results,
in which the coefficients grow much less than factorially.
We apply our formula to Drell--Yan pair production, heavy flavour production,
and the production of high invariant mass jet pairs in hadronic collisions.
We find that, with our formula, resummation effects become important
only fairly close to the threshold region.
In the case of heavy flavour production
we find that resummation effects are small
in the experimental configurations of practical interest.
}
\end{abstract}
\vfill
\noindent
CERN-TH/96-86 \newline
March 1996    \hfill
\end{titlepage}

\section{Introduction}
In this work we deal with the problem of the resummation of logarithmically
enhanced effects  in the vicinity of
the threshold region in hard hadroproduction processes.
Drell--Yan lepton pair production has been in the past the best
studied example of this sort \cite{oldresummrefs,AEM,VanNeervenDY}.
The threshold region
is reached when the invariant mass of the lepton pair
approaches the total available energy.
A large amount of theoretical
and phenomenological work has been done on this subject.
References \cite{Sterman} and \cite{CataniTrentadue} summarize all the
theoretical progress performed in this field.
Resummation formulae have also been used in estimating heavy flavour
production \cite{Laenen}, \cite{BergerContopanagos}. In this case only a
leading logarithmic resummation formula is known.
Calculations of the next-to-leading logarithms are in progress
\cite{StermanLathuile}.

In the present work, we will mostly be concerned with difficulties
that arise when one tries to apply resummed formulae to physical processes.
This is a highly non-trivial problem. What one
typically finds is that resummation involves the integration of the
running coupling over the Landau pole, which has to be regulated.
In early works on resummation in the Drell--Yan cross section, the problem
was avoided by either assuming a fixed coupling constant (Curci and Greco
in ref.~\cite{oldresummrefs}), or
by shifting the argument of $\as$ so as to move the position of
the Landau pole to $Q^2=0$ (Parisi and Petronzio in ref.~\cite{oldresummrefs}).
In ref. \cite{Appel} a cut-off procedure was introduced in order to regulate
these singularities.
A similar approach was used in ref.~\cite{Laenen}, in the context of
heavy flavour hadroproduction.
In refs. \cite{StermanContopanagos} and \cite{AlveroContopanagos}
a principal value
prescription was adopted instead, and an application to top production
was proposed in ref.~\cite{BergerContopanagos}.
It was generally found that threshold corrections become quite large,
long before the hadronic threshold region is reached.

Lately, the problem of the presence of an integration over the
Landau pole in resummation formulae has been
reexamined from the point of view of the occurrence of infrared renormalons
(IR) in the QCD perturbative expansion for the Drell--Yan process
\cite{StermanContopanagos,KorchemskiSterman,AkhouryZacharov1}.
Roughly speaking, the IR point of view relates the factorial growth of the
coefficients of the perturbative expansion to the presence of
power-suppressed corrections to the process in question. It is found that the
ambiguities associated to the resummation of the non-convergent (asymptotic)
perturbative expansion have precisely the form of a power-suppressed
correction. In this context the separation of perturbative
and non-perturbative effects is at best ambiguous, since it relies on
a specific prescription needed to resum an asymptotic expansion.
In refs.~\cite{StermanContopanagos,KorchemskiSterman,AkhouryZacharov1}
it was argued that the Landau pole integration in the resummation
formulae for threshold corrections gives the leading IR behaviour of the
perturbative expansion of the Drell--Yan cross section, and that the
associated factorial growth is the one corresponding to a $1/Q$
ambiguity. A recent work of Beneke and Braun \cite{BenekeBraun}
has however demonstrated that the approximations made in the resummation
formulae for the logarithmic corrections are insufficient to correctly
describe the IR structure,
and that when higher order contributions are properly included
no factorially growing
terms associated to a $1/Q$ ambiguity do arise in the Drell--Yan cross section.
This result was subsequently confirmed in refs.~\cite{WebberDokshitzer}
and \cite{AkhouryZacharov2}. In the latter reference,
the absence of $1/Q$ effects was shown to be a consequence of cancellations
related to the Kinoshita-Lee-Nauenberg theorem.

The IR point of view teaches us a very important lesson
to keep in mind: it is not enough to make
sure that all the leading corrections are properly included  in
the perturbative expansion. We should also make sure that formally
subleading terms, which are not controlled in our approximation,
will not affect the asymptotic property of the expansion.
In fact, if formally subleading terms happen to have a strong factorial
growth, they may induce large corrections even in kinematic regions
where our resummation is not justified.

In the present work we find that besides the IR problem, other,
more important sources of factorial growth may be introduced,
which are spurious and are by no means implied by the
threshold approximation, since they are not enhanced at threshold.
These large terms arise when one attempts to formulate the resummation
problem in $x$-space, as opposed to its natural formulation in moment
(or $N$) space. To be specific, let us consider the case
of the Drell--Yan pair production. The Drell--Yan
cross section can be written in $x$ space (schematically) as
\beq
\sigma^{\rm (DY)}(\tau)=\int_0^1 dx_1\;dx_2\;dx\;F(x_1)\,F(x_2)\;
\hat\sigma^{\rm (DY)}(x) \delta(x_1\,x_2\,x-\tau) \; .
\eeq
In moment space a simple factorized expression follows
\beq
\sigma^{\rm (DY)}_N=F_N\;F_N\;\hat\sigma^{\rm (DY)}_N \; ,
\eeq
where
\beqn
F_N&=&\int_0^1 \frac{dx}{x}\;x^N\; F(x)
\nonumber\\
\sigma^{\rm (DY)}_N&=&\int_0^1 \frac{d\tau}{\tau}\;\tau^N\;
\sigma^{\rm (DY)}(\tau)
\nonumber\\
\hat\sigma^{\rm (DY)}_N&=&\int_0^1 \frac{dx}{x}\;x^N\;
\hat\sigma^{\rm (DY)}(x)\;.
\eeqn
The threshold region $\tau \to 1$ corresponds to the limit $N \to \infty$ in
$N$-space. In this limit, soft-gluon radiation produces large logarithmic
corrections of the type $\as^n \ln^m N$ that are resummed in the partonic cross
section $\hat \sigma$.
Resummation of soft-gluon effects is best expressed in $N$-moment space,
because it leads to the exponentiation of the logarithmic corrections.
Exponentiation is a consequence of dynamics and kinematics factorization.
By dynamics we mean factorization of multigluon QCD amplitudes to logarithmic
accuracy. By kinematics we mean factorization of the phase space:
the constraint of longitudinal-momentum conservation factorizes in
$N$-moment space.

However, the moment space formula can be turned to an $x$-space
formula. We will see that with this transformation,
by neglecting certain subleading terms,
one may generate large, factorially growing corrections,
which may be wrongfully attributed
to the original resummation formula. In fact these subleading terms
are there to compensate for the fact
that exponentiation is imperfect in the $x$-space formulation,
and should not be dropped.
If they are neglected, kinematic constraints that were satisfied
in the original formulae (e.g. momentum sum rules) are strongly
violated, and actually diverge,
in the $x$-space expression. A typical consequence of this
procedure is that the final formula for the physical cross section receives
large soft gluon corrections (actually, divergent ones)
even if we are far from the threshold region.
With these large factorially growing terms, a
corresponding power corrections of the order of $(\Lambda_{\rm QCD}/Q)^\delta$
can be associated,
where $\delta$ can be much less than 1. The usual ``common sense''
assumption that power corrections arise from regions of phase
space where the momenta are of order $\Lambda_{\rm QCD}$
seems therefore to fail, and one is forced to use a cut-off
of several GeV
in order to make any sense out of the resulting formulae.
We argue that all these paradoxes are simply avoided if the
transformation to $x$-space is performed exactly.

In the present work we provide a specific prescription for the
resummation of soft gluon effects that does not have any factorially
growing terms in its perturbative expansion. The ambiguity associated
with the perturbative expansion of our formula is therefore free of any
$1/Q^M$ effects for any $M\ge 1$, and in fact turns out to be of the
form $e^{-H\,Q(1-\tau)/\Lambda}$, where $\tau$ is the ratio of the
squared invariant mass $Q^2$ of the produced heavy system and of the
total CM energy squared, $H$ is a slowly varying positive function,
and $\Lambda$ is the QCD scale parameter.

The fact that the ambiguities introduced by the Landau pole only arise
for values of $1-\tau$ of the order of $\Lambda/Q$ is consistent with
our expectation that a correct resummation of soft logarithms should
allow control over the perturbative expansion down to the scale at
which the coupling constant blows up, namely $\Lambda$.  A successful
resummation program, in fact, should work regardless of the size of
the logarithmic terms being resummed, provided one can prove that the
neglected terms are sufficiently small. The non-perturbative regime is
not defined by the region of momenta in which higher order terms are
larger than lower order ones; it is defined by the domain in which
the terms which are neglected by the resummation procedure are
comparable in size with those taken into account.  The universality of
soft gluon emission should allow full control to be maintained over the
dominant contributions to the perturbative expansion even when they
become large, and to correctly resum them regardless of their size.
Our approach to resummation shows that this is indeed possible,
confining the effects of the really non-perturbative regime of QCD to
their natural scale, namely $\Lambda$.

The paper is organized as follows.
In section~2 we will give a few reference formulae and establish
our notation. We will mainly deal with the Drell--Yan case as an illustrative
example. In section~3 we will show how large spurious corrections
may arise in the computation of the resummed Drell--Yan cross sections.
In section~4 we will propose an alternative resummation method, and
in section~5 we discuss its implications for Drell--Yan cross sections.
In sections~6 and 7 we discuss heavy flavour production, and
in section~8 jet production
at large transverse momenta.
In section~9 we discuss a few remaining issues, and
in section~10 we give our conclusions.
In Appendix~A we prove the absence of factorially growing terms
in our resummation formula, in Appendix~B we discuss some
details about the numerical method we used, and in Appendix~C we derive
some results regarding the inverse Mellin transform in the leading logarithmic
approximation.

\section{Basic formulae and notation}

We begin with the formula for Drell--Yan pair production
\beq \label{DYCrossSection}
\sigma(\tau, Q^2)=\int_0^1 dx\,dx_1\,dx_2 F(x_1) F(x_2)
\delta(x x_1 x_2 - \tau)
\Delta(x,Q^2)\,.
\eeq
At the Born level, omitting obvious factors,
we have $\Delta(x,Q^2)=\delta(1-x)$.
We also omit, for ease of notation,
the parton indices. The cross section is given in terms of the parton
densities $F(x)$ as measured in the deep inelastic processes.
Defining the Mellin transform as
\beq
f_N=\int_0^1 \frac{dz}{z}\,z^N\,f(z)
\eeq
we can rewrite eq.~(\ref{DYCrossSection}) as
\beq
\sigma_N(Q^2)=F_N^2(Q^2)\,\Delta_N(Q^2)\,.
\eeq
The resummed coefficient function for the Drell--Yan
process in the DIS scheme is
\cite{oldresummrefs,Sterman,CataniTrentadue}
\beqn\label{deltan}
\ln \Delta_N(Q^2)&=& - \int_0^1 dx \;\frac{x^N -1}{1-x} \;
\left[ 2 \int_{(1-x)^2Q^2}^{(1-x)Q^2}
\frac{dq^2}{q^2} A(\as(q^2))+  B(\as((1-x)Q^2)) \right]\nonumber \\
&+& {\cal O}(\as(\as \ln N)^k)\;\;,
\eeqn
with
\beq\label{AB}
A(\as)={\asp} A^{(1)}+\left(\asp \right)^2 A^{(2)}\;,\;\;\;\;\;\;
B(\as)={\asp} B^{(1)}
\eeq
where ($ C_A = 3 \,, \;\;C_F = 4/3\,, \;\;T_R = 1/2 \;\;$ in QCD)
\beq\label{AB12}
A^{(1)}= C_F\;,\;\;\;\; A^{(2)}=\frac{1}{2} C_F K \;,\;\;\;\;\;
B^{(1)}=- \frac{3}{2} C_F \;\;,
\eeq
and the coefficient $K$ is given by \cite{Kodaira}
\beq\label{kcoef}
K = C_A \left( \frac{67}{18} - \frac{\pi^2}{6} \right)
- \frac{10}{9} T_R N_f \;.
\eeq
Note that, due to the integration of the running coupling, the integral
in eq.~(\ref{deltan}) is singular for all
values of $N$. However, if we perform the integration to next-to-leading
logarithmic (NLL) accuracy (i.e. we compute the leading $\as^n \ln^{n+1}N$ and
next-to-leading $\as^n \ln^{n}N$ terms)
\cite{CataniTrentadue}, we find
\beq\label{lnd}
\ln \Delta_N(Q^2) = \ln N \;g_1(\bzero\as\ln N)
+ g_2(\bzero\as\ln N) + {\cal O}(\as^k \ln^{k-1}N)
\eeq
where $\as = \as(Q^2)$, $\bzero$ and $\bone$ are the first two coefficients of
the QCD $\beta$-function
\beq\label{betas}
\bzero = \frac{11 C_A - 4 T_R N_f}{12\pi}\;,\;\;\;\;\; \bone =
\frac{17 C_A^2 - 10 C_A T_R N_f -6 C_F T_R N_f}{24\pi^2}\;,
\eeq
and the leading and next-to-leading functions
$g_1$ and $g_2$ are given by ($\GE = 0.5772\ldots$ is the Euler constant)
\beqn\label{g0}
g_1(\lambda) =
&+&\frac{A^{(1)}}{\pi\bzero\lambda}\Bigl[(1-2\lambda)
\ln(1-2\lambda)-2(1-\lambda)
\ln(1-\lambda)\Bigr] \; , \\
\label{g1}
g_2(\lambda) = &+&\frac{A^{(2)}}{\pi^2\bzero^2}\Bigl[2\ln(1-\lambda)
-\ln(1-2\lambda)\Bigr] \nonumber \\
&-&\frac{B^{(1)}}{\pi\bzero}\ln(1-\lambda) +\frac{2
  A^{(1)}\GE}{\pi\bzero}\Bigl[\ln(1-\lambda)
-\ln(1-2\lambda)\Bigr] \\
&+&\frac{A^{(1)}\bone}{\pi\bzero^3}\left[\ln(1-2\lambda)
  -2\ln(1-\lambda)+\half\ln^2(1-2\lambda)-\ln^2(1-\lambda)\right]\; .
\nonumber \eeqn
We observe that the $N$-space formula eq.~(\ref{lnd}) is
finite and uniquely defined up to the very large value of
$N=N_L=\exp\frac{1}{2\as\bzero}$, in spite of the fact that it is obtained from
an expression which is formally divergent for any value of $N$. It was shown
in ref.~\cite{CataniTrentadue} that the
expression $x^N-1$ in formula~(\ref{deltan}) can be replaced to NLL accuracy
by the theta function $-\theta(1-x-e^{-\gamma_E}/N)$.
The region where the integral is
divergent is therefore excluded from the integration if $N < N_L$.
This shows that the
divergences present in the integral of eq.~(\ref{deltan}) are subleading
for large $N$. They
may be cancelled by other divergences of the same nature, neglected by the
approximations that lead to formula~(\ref{deltan}). This is indeed the case
for the leading $1/Q$ singularity, as shown in ref.~\cite{BenekeBraun}.

Our result in eq.~(\ref{lnd}) can be expressed in terms of $\as(\mu^2)$ for an
arbitrary value of the renormalization scale $\mu^2$. We thus achieve full
control over the {\em renormalization-scale} dependence. To do that, we must
take into account the scale dependence of the next-to-leading
function $g_2$, which is given by
\beq\label{g2scale}
g_2(\lambda,\mu^2) = g_2(\lambda,Q^2)
+\lambda^2 g'_1(\lambda)\ln(\mu^2/Q^2)\;,
\eeq
where $g_2(\lambda,Q^2)$ is $g_2(\lambda)$ as defined above and
\beq\label{g1pr}
g'_1(\lambda) = - \frac{A^{(1)}}{\pi\bzero\lambda^2}\Bigl[\ln(1-2\lambda)
-2\ln(1-\lambda)\Bigr]\; .
\eeq

Also the {\em factorization-scheme} dependence is completely under
control up to NLL accuracy~\cite{CMW}.
In the ${\overline {\rm MS}}$ scheme the resummed
coefficient function can still
be expanded as in Eq.~(\ref{lnd}), that is:
\beq\label{lndms}
\ln \Delta_N^{{\overline {\rm MS}}}(Q^2) = \ln N
\;g_1^{{\overline {\rm MS}}}(\bzero\as\ln N)
+ g_2^{{\overline {\rm MS}}}(\bzero\as\ln N) + {\cal O}(\as^k \ln^{k-1}N) \;\;,
\eeq
and the leading and next-to-leading functions
$g_1^{{\overline {\rm MS}}}$ and $g_2^{{\overline {\rm MS}}}$ are given by
\beq\label{g0ms}
g_1^{{\overline {\rm MS}}}(\lambda) =
+\frac{A^{(1)}}{\pi\bzero\lambda}\Bigl[ \,2\lambda + (1-2\lambda)
\ln(1-2\lambda) \Bigr] \; ,
\eeq
\beqn\label{g1ms}
g_2^{\overline {\rm MS}}(\lambda) =
&-& \frac{A^{(2)}}{\pi^2\bzero^2}\Bigl[2\lambda
+\ln(1-2\lambda)\Bigr] - \frac{2 A^{(1)}\GE}{\pi\bzero} \;\ln(1-2\lambda)
\nonumber \\
&+&\frac{A^{(1)}\bone}{\pi\bzero^3}\left[ 2 \lambda + \ln(1-2\lambda)
+\half\ln^2(1-2\lambda) \right] \;\; .
\eeqn
In this scheme, the renormalization-scale dependence is again taken
into account by the function $g_2$ as in eq.~(\ref{g2scale}), and the
corresponding function $g'_1$ is:
\beq
g_1^{\prime \,{\overline {\rm MS}}}(\lambda) = - \frac{A^{(1)}}{\pi\bzero\lambda^2}
\Bigl[\ln(1-2\lambda) +2 \lambda \Bigr]\; .
\eeq

Sometimes, as an illustration, we will also use the double log
approximations (DLA) to the resummation formulae
\beqn
g_1(\lambda)&=&\frac{A^{(1)}}{\bzero\pi}\lambda+{\cal O}(\lambda^2)\;,
\\
g_1^{{\overline {\rm MS}}}(\lambda) &=&
 \frac{2\, A^{(1)}}{\pi\bzero}\,\lambda+{\cal O}(\lambda^2)\;,
\eeqn
which give rise to the expressions
\beqn
\ln \Delta_N(Q^2)&=&\frac{A^{(1)}}{\pi}
\as\ln^2 N+{\cal O}((\as\ln N)^{k+1}\ln N)\;,
\\
\ln \Delta^{\overline {\rm MS}}_N(Q^2)&=&\frac{2\,A^{(1)}}{\pi} \as \ln^2 N
+{\cal O}((\as\ln N)^{k+1}\ln N)\;.
\eeqn
\section{Problems with $x$-space resummation formulae}\label{Problems}
In this section we discuss the problems that
may arise when turning the $N$-space resummation formula into an
$x$-space formula.
We begin by considering the DLA case.
No running coupling effect is present in this
case. Therefore, there is no Landau pole, and the result of resummation
should be finite and free of ambiguities. There is a ``realistic''
limit that corresponds to this case, which is the limit of large colour
factors and small coupling. In the case of gluon initiated processes
in the \MSB\ scheme, this limit is not far from reality.
For the sake of definiteness, we also fix the structure function
\beq
F_N=\frac{6}{N(N+1)(N+2)}
\eeq
which corresponds to the $x$-space structure function
\beq
F(x)=(1-x)^2\;.
\eeq
Our cross section
formula is
\beq\label{xsecx}
\sigma(\tau)=\frac{1}{2\pi i}
\int_{C-i\infty}^{C+i\infty}\; F^2_N\,\Delta_N\;\tau^{-N}\; dN\;,
\eeq
and we now take the simplified soft gluon factor to be given by
\beq
\Delta_N=\exp(\,a\log^2 N\,)\;,
\eeq
where only the double logarithmic term has been kept in $\Delta_N$.
The coefficient $a$ is both process and scheme dependent
(see Sects.~\ref{hfp} and \ref{jcs});
it is given by $a=C_F\as/\pi$ and $a=2 C_F\as/\pi$ for the $q\bar{q}$
initial state in the DIS and \MSB\ scheme respectively, and
by $a=2 C_A\as/\pi$ for the $gg$ initial state in the \MSB\ scheme.
The integral (\ref{xsecx}) is not absolutely convergent for large $N$,
since $\Delta_N$
grows faster than any power for large $N$. Observe, however,
that if we expand $\Delta_N$
in powers of $a$, the integral converges order by order in perturbation
theory. In the corresponding perturbative expansion, we can therefore
deform the integration contour into two straight half-lines
from $C-(i+\ep)\infty$ to $C$, and then to $C+(i-\ep)\infty$.
Once the perturbative expansion is written in this way we realize
that it can be resummed into the expression
\beq\label{xsecx_ep}
\sigma(\tau)=\frac{1}{2\pi i}\int_{C-(i+\ep)\infty}^{C+(i-\ep)\infty}\;
F^2_N\,\Delta_N\;\tau^{-N}\; dN\;,
\eeq
which is now convergent, and independent of $\ep$.
In the following, we will always interpret the integral in the above
sense, omitting the explicit reference to the $\ep$.

Let us now see what happens if we try to rewrite eq.~(\ref{xsecx}) as an
$x$-space formula. First of all we need the inverse Mellin transform
of $\Delta_N$. We have (see Appendix C and ref.~\cite{CataniWebber})
\beqn
\Delta(x)&=&\frac{1}{2\pi i}
\int_{C-i\infty}^{C+i\infty}\;\exp(a\log^2 N)\;x^{-N}\; dN\;
\nonumber \\ \label{dlantodlax}
&=&-\frac{d}{d x}\left( \theta(1-\eta-x)\exp[a\log^2(1-x)]\right)
\times (1+{\rm NLL\;\;terms})
\eeqn
where, according to the usual definitions, NLL terms
stand for contributions of the form
$\as^k \log^m 1/(1-x)$ with $k\ge 1$ and $m\le k$.
The r\^ole of the $\theta$ function is to define the right-hand side in
a distribution sense as $\eta\to 0$, so that we have the correct normalization
of the first moment
\beq
\int_0^1 \Delta(x)\;dx\;=\;1\;.
\eeq
Thus, defining
\beq
{\cal L}(z)=\int_z^1
 \frac{dx_2}{x_2}\;F\left(\frac{z}{x_2}\right)\;F(x_2)
\eeq
and neglecting NLL terms, we obtain the following expression
for the cross section
\beqn
\sigma(\tau)&=&\int_0^1 dx\,dx_1\,dx_2\, F(x_1)\,F(x_2)
\delta(x\,x_1\,x_2-\tau)\,\Delta(x)\nonumber \\
&=&\int_\tau^1\,dx\,\exp[a\log^2(1-x)]\frac{d}{dx}
{\cal L}\left(\frac{\tau}{x}\right)\;,
\label{sigmadoublelog}
\eeqn
where an integration by part was performed. We now see that the integral
in eq.~(\ref{sigmadoublelog}) is divergent at $x=1$ for any value of
$\tau$, since the expression $\exp[a\log^2(1-x)]$ diverges faster than
any power as $x\to 1$. Let us examine more carefully the origin of this
divergence. If we expand eq.~(\ref{sigmadoublelog}) in powers of $a$, each term
of the expansion is integrable, but the corresponding series is divergent.
Since the term $d/dx\;{\cal L}(\tau/x)$ is a smooth function of
$x$ as $x\to 1$,
the nature of the divergence is given by the following integral
\beq\label{divergentnature}
\int_0^1 \exp[a\log^2(1-x)]\;dx
=\sum_{k=0}^\infty\; \frac{a^k}{k!}\;\int_0^1 \log^{2k}z\;dz
=\sum_{k=0}^\infty\; \frac{a^k(2k)!}{k!}\;.
\eeq
The asymptotic behaviour of the coefficients for large $k$
is $(2k)!/k!\approx 4^k\,k!$. The expansion
is therefore an asymptotic one. Observe that the lower limit of the
integral in eq.~(\ref{divergentnature}) is irrelevant for this conclusion.
It is known that factorially growing terms in the perturbative expansion
are associated to power-like ambiguities in the resummed expression.
In order to resum the asymptotic expansion, we should in fact truncate the
series when the next term is of the same size as the current one, i.e.
when $4ak=1$. The error on the resummed expression is then of the order
of the left over term
\beq
\delta=(4a)^k\,k!\approx (4a)^k\,k^k\,e^{-k}=e^{-\frac{1}{4a}}\,.
\eeq
If we replace the appropriate value of $a$ we get
\beqn
\delta&=&\left(\frac{\Lambda}{Q}\right)^{\frac{\pi\bzero}{2\,C_F}}
\quad\quad q\bar{q},\quad {\rm DIS},
\\
\delta&=&\left(\frac{\Lambda}{Q}\right)^{\frac{\pi\bzero}{4\,C_F}}
\quad\quad q\bar{q},\quad \MSB,
\\
\delta&=&\left(\frac{\Lambda}{Q}\right)^{\frac{\pi\bzero}{4\,C_A}}
\quad\quad gg,\quad \MSB.
\eeqn
Although it is power-suppressed, the smallness of the exponent makes this
effect potentially large.
Thus, for example, for $\nf=5$ we have for
Drell--Yan in the DIS scheme a $(\Lambda/Q)^{0.72}$ power correction,
and for heavy flavour production via gluon fusion at fixed invariant
mass of the heavy flavoured pair we have
a $(\Lambda/Q)^{0.16}$ correction, which is hardly distinguishable
from a correction of order~1.

Instead of truncating the perturbative expansion, we may achieve the
same goal by putting a cut-off in the integral. In fact, consider
the cut-off integral
\beq
\int_0^{x_0} dx\;\log^{2k}\frac{1}{1-x}=
\int_0^{\log\frac{1}{1-x_0}} dt\;t^{2k}\;e^{-t}\;.
\eeq
The saddle point of the integral is at $t=2k$. If the saddle point
is within the integration range, the integral is essentially the
factorial of $2k$, while for larger values of $k$ it starts to
grow like a simple power. Therefore, the cut-off acts like
a truncation of the expansion. In order to have, as before, a truncation
at $k=1/(4a)$, we need to set the cut-off at $\log 1/(1-x_0)=2k=1/(2a)$,
corresponding to
\beq\label{cutoff}
1-x_0=e^{-\frac{1}{2a}}\;.
\eeq
In the worst case of production via gluon fusion in the \MSB\ scheme,
we would have $1-x_0=(\Lambda/Q)^{0.32}$. This leads to the rather
paradoxical conclusion that the cut-off on the soft radiation
should be imposed at values of $Q$ much larger than $\Lambda$.
For example, for the production of a 100 GeV object we would need
a cut-off of the order of 14 GeV. This would have to increase
with energy. We observe that cut-offs of this kind are used in
ref.~\cite{BergerContopanagos}, in formulae (114) and (115).
As a matter of fact, if we take the limit of large colour factor and
small coupling
of the formulae given there, we recover eq.~(\ref{cutoff}).

\begin{figure}[htb]
\centerline{\psfig{figure=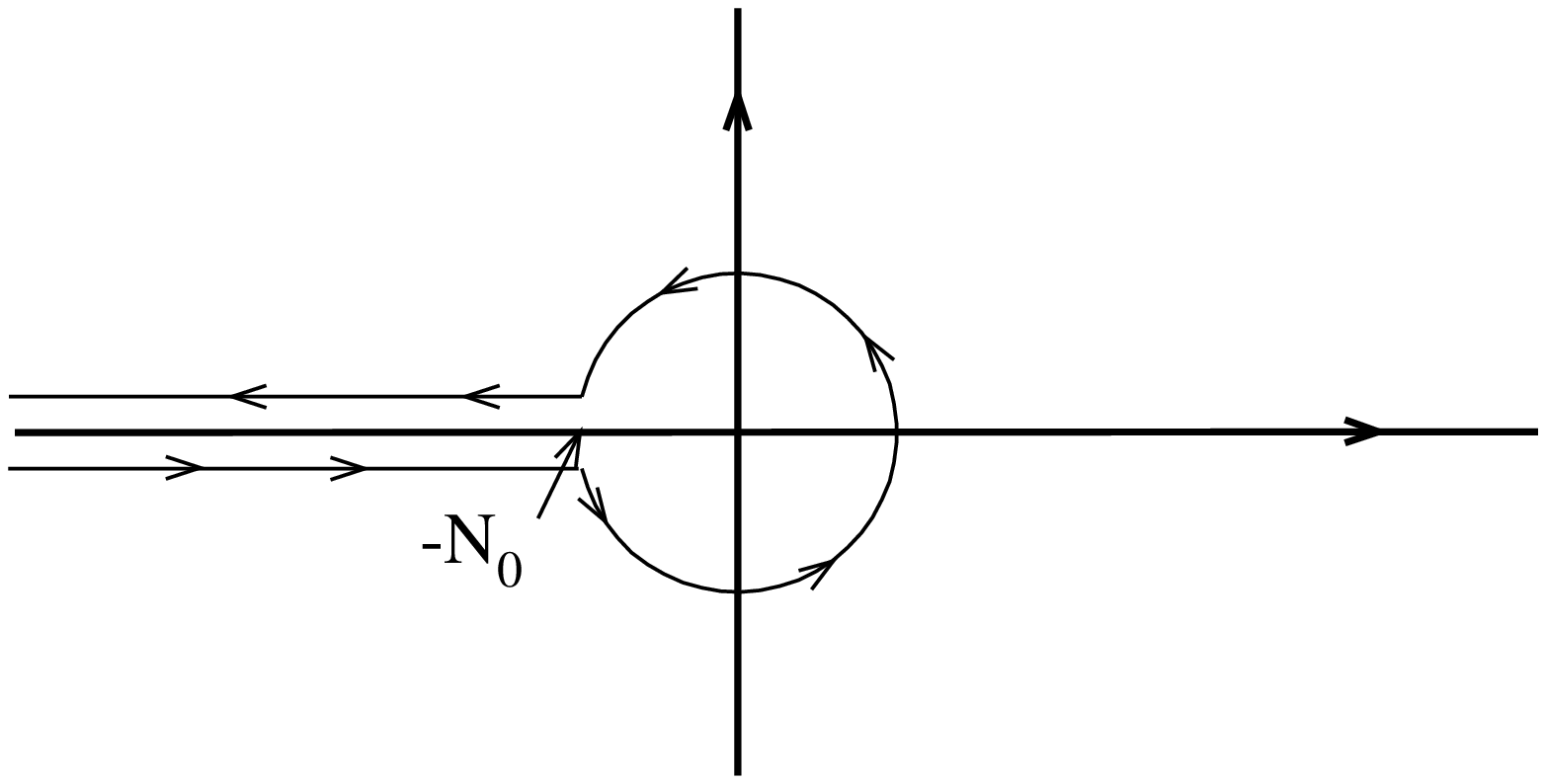,width=10cm,clip=}}
\ccaption{}{ \label{contour1}
Integration contour for $N$ for the determination of asymptotic properties.}
\end{figure}

From the above derivation we see that the large corrections obtained
have nothing to do
with infrared renormalons, and it is easy to convince ourselves that they
are a spurious effect.
They were in fact not present in the original expression,
eq.~(\ref{xsecx}), which is finite. It is also easy to show that
the perturbative expansion of eq.~(\ref{xsecx}) has no factorially growing
terms. This can be done in the following way.
The $k^{\rm th}$ coefficient of the expansion is given by the integral
\beq
c_k=\frac{1}{2\pi i}\int_{C-i\infty}^{C+i\infty}
\frac{36}{(N(N+1)(N+2))^2}\frac{1}{k!}\log^{2k}N\;\;\tau^{-N} \;dN\,.
\eeq
We deform the integration
contour as illustrated in fig.~\ref{contour1}.
By choosing $N_0$ sufficiently large, we get two contributions to our
integral, from the circle and from the discontinuity along the negative
axis. The integral on the circle is bounded by the expression
$C \log^{2k} N_0\;/k!$ for some value of $C$. The discontinuity integral
is instead given by (replacing $N\to -N$)
\beqn
&& \frac{36}{k!}\int_{N_0}^\infty
\exp[N\log\tau-2\log(N(N-1)
(N-2))]\;{\rm Disc}[\log^{2k}(-N)]\;dN
\nonumber\\ && \leq
 \frac{36}{k!}\int_{N_0}^\infty
\exp[N\log\tau+2k\log(\pi+\log N)]\;dN\,.
\eeqn
By saddle point integration of the right hand side
we immediately see that the above expression cannot grow faster than
\beq
\frac{1}{k!}\;\left(\frac{\log 2k}{\log 1/\tau}\right)^{2k}\,.
\eeq
Therefore the power expansion of eq.~(\ref{xsecx}) has an infinite radius of
convergence, and, {\it a fortiori}, does not have factorially growing terms.

Apparently, when performing the inverse Mellin
transform to obtain the $x$-space expression of the cross section,
we have simply thrown away subleading
terms that would have compensated the factorial growth.
In a leading-log sense we have every right to throw away subleading terms.
However, if by doing so we generate unjustified factorially growing
terms, we are certainly doing something wrong on physical ground,
even if we are perfectly consistent with the leading-log approximation.

In the case of the full resummation formula, including the effects of the
running coupling, the above illustrated problem persists.
We have in this case (see Appendix~C)
\beqn
\Delta(x)&=&\frac{1}{2\pi i}
\int_{C-i\infty}^{C+i\infty}\;
\exp(\log N \,g_1(\as\bzero\log N))\;x^{-N}\; dN\;
\nonumber \\ \label{dlantodlax1}
&=&-\frac{d}{d x}\left( \theta(1-\eta-x)\exp[l \, g_1(\as\bzero l)]\right)
\times (1+{\rm NLL\;\;terms})
\eeqn
in which an unintegrable singularity is met before $x$ reaches 1,
the Landau pole singularity.
In the corresponding formula for the partonic cross section, neglecting
NLL terms, we get
\beq
\sigma(\tau)=\int_\tau^1\,dx\,\exp[l \, g_1(\as\bzero l)]\frac{d}{dx}
{\cal L}\left(\frac{\tau}{x}\right)\;.
\eeq
Expanding the above formula in powers of
$\as$, we would generate the same type of spurious factorial growth
as found before. As before, the nature of the divergence is given by the
integral
\beq \label{divergentnature1}
\int_0^1\,dx\,\exp[l \, g_1(\as\bzero l)]\;.
\eeq
Using commonly available algebraic programs, it is easy to expand
eq.~(\ref{divergentnature1}) up to large orders, and then study
numerically the factorial growth. Expanding up to $\as^{32}$
we have found the behaviour $k! C_{(k)}^k (b_0\as)^k$, where
$C_{(k)}$ is a slowly increasing function of $k$. If, for large $k$,
$C_{(k)}$ approaches a limiting value $C$, this corresponds
to a power ambiguity of $(\Lambda/Q)^{2/C}$. For gluon
fusion in the \MSB\ scheme we get $C_{(32)}=10.48$,
corresponding to a power ambiguity of $(\Lambda/Q)^{0.19}$, while for
$q\bar{q}$ annihilation in the \MSB\ scheme we get $C_{(32)}=4.0$,
corresponding
to $(\Lambda/Q)^{0.5}$. These numbers are roughly consistent with
those of the exact analysis performed for the fixed coupling case.

We observe that, even if we modify the above $x$-space formula,
by expanding it in powers of $\as$ and keeping only a fixed number
of terms, the problem discussed earlier still persists. In fact, our discussion
is relative to the case in which the exponent in formula
(\ref{dlantodlax1}) is expanded and truncated to order $\as$.

An $x$-space resummation procedure, similar to the one discussed here,
has indeed been adopted in the literature.
In refs.~\cite{Laenen,KidonakisSmith95}
a cut-off procedure is applied to screen the Landau pole
singularity that manifests itself when $x$ approaches 1.
This cuts off both the divergence
due to the Landau pole, and the spurious divergence we just described.
We therefore argue that the uncertainties induced by this procedure
are much larger than needed, since they introduce a divergence that
is in fact not present.
In ref.~\cite{BergerContopanagos}, the Landau pole singularity is
dealt with by using a principal value prescription.
Subsequently, a cut-off is introduced in order to screen
the large subleading effects that arise when performing the Mellin
transform of the partonic cross section from $N$ to $x$ space.
Our point is precisely that if these large subleading terms had
been kept, they would have cancelled the factorially growing terms
arising from the leading terms after integration against the parton
luminosities.
In all these approaches,
unphysically large cutoffs are needed in order to avoid
the large corrections that arise at higher order in the perturbative expansion.

We conclude that, when proposing a resummation formula for threshold effects,
it is not enough to make sure that all leading terms are included in the
formula. We must also make sure that we are not introducing subleading terms
that grow very fast with the order of the perturbative expansion in the final
physical result.
In the next section we propose a resummation formula that is correct
from the point of view of the threshold approximation, but does not
induce any factorial growth in the perturbative expansion.

\section{The Minimal Prescription formula}
Our starting point is eq.~(\ref{deltan}).
Its perturbative expansion has the form
\beq \Delta_N=\sum_{k=0}^\infty\;c_k(\log N)\,\as^k \eeq
where the coefficients $c_k(\log N)$ are polynomials in $\log N$.  The
resummed cross section can be formally written as a power expansion
\beq\label{MPexp}
\sigma_\tau=\frac{1}{2\pi i}\sum_{k=0}^\infty\;\as^k\;
\int_{C-i\infty}^{C+i\infty} F^2_N(Q^2)\;c_k(\log N) \;\tau^{-N}\;dN\;.
\eeq
Observe that the integrals in the coefficients of the expansion are finite for
$C>2$, and no singularities occur at the right of the integration contour.
The choice of $C>2$ is motivated by the usual Regge behaviour of structure
functions, which
implies that we cannot have any singularity in $F_N(Q^2)$ to the
right of the pomeron singularity, which is slightly above $N=1$.
The Landau pole for $N=\exp\frac{1}{2\as\bzero}$ manifests itself
in the fact that the series~(\ref{MPexp}) is not convergent.

We now propose the following formula for the resummation of
threshold effects in the Drell--Yan cross section
\beq\label{MP}
\sigma_{\rm res}(\tau)=\frac{1}{2\pi i}
\int_{C_{\rm MP}-i\infty}^{C_{\rm MP}+i\infty}\;
F^2_N(Q^2)\,\Delta_N(Q^2)\;\tau^{-N}\; dN\;,
\quad 2<C_{\rm MP}<N_L\equiv\exp\frac{1}{2\as \bzero}\;,
\eeq
where $\Delta_N(Q^2)$ is given in eq.~(\ref{lnd}).
The constant $C_{\rm MP}$ is chosen in such a way that all singularities
in the integrand are to the left of the integration contour,
except for the Landau singularities
at $N=N_L$ and $N=N_L^2$,
which lie to the far right.
We will call eq.~(\ref{MP}) the ``Minimal Prescription'' (MP) in the following.
Its justification relies on the following
important properties, which will be proved in Appendix~A:
\begin{itemize}
\item
The expansion (\ref{MPexp}) converges asymptotically to the MP formula.
Observe that this would not happen if we had chosen a contour that
passes to the right of the first Landau pole.
\item
The coefficients of the expansion (\ref{MPexp}) do not grow factorially.
\item
If we truncate the expansion (\ref{MPexp}) at the order
at which its terms are at a minimum, the difference between
the truncated expansion and the full MP formula is suppressed by a factor
\beq
e^{-H\frac{Q(1-\tau)}{\Lambda}}\,,
\eeq
where $H$ is a slowly varying positive function. This suppression factor
is stronger than any power suppression.
\end{itemize}
We stress that with our MP formula we do not introduce any spurious factorial
growth in the perturbative expansion. One may object that in this way
we do not introduce any possible renormalon effect in the formula.
Factorial growth due to renormalons is very likely to be present
in the perturbative expansion. Our point is, however, that the leading
terms in our expansion do not necessarily contain this factorial growth,
and that renormalons present in a resummed expression therefore
do not necessarily reflect the renormalons present in the full perturbative
expansion. To be more specific, let us consider for a moment
eq.~(\ref{deltan}). It is clear that, if we perform the $x$ integration
exactly, we are indeed integrating over the Landau pole. However, since this
formula is accurate at the NLL level at most, we may integrate it
in the NLL approximation, and obtain formula (\ref{lnd}), which has no trace of
factorial growth.
In particular, it was shown in ref.~\cite{BenekeBraun}
that the leading IR arising from naively extending Eq.~(\ref{deltan})
beyond the NLL level
cancel in the full perturbative expansion. We take this result as a
confirmation of the fact that the resummation of logarithmic effects
at threshold does
not teach us anything about the structure of power corrections.
Resummation formulae should not, therefore, include any
power correction.
\section{The Drell--Yan cross section}
\begin{figure}[htb]
\centerline{\psfig{figure=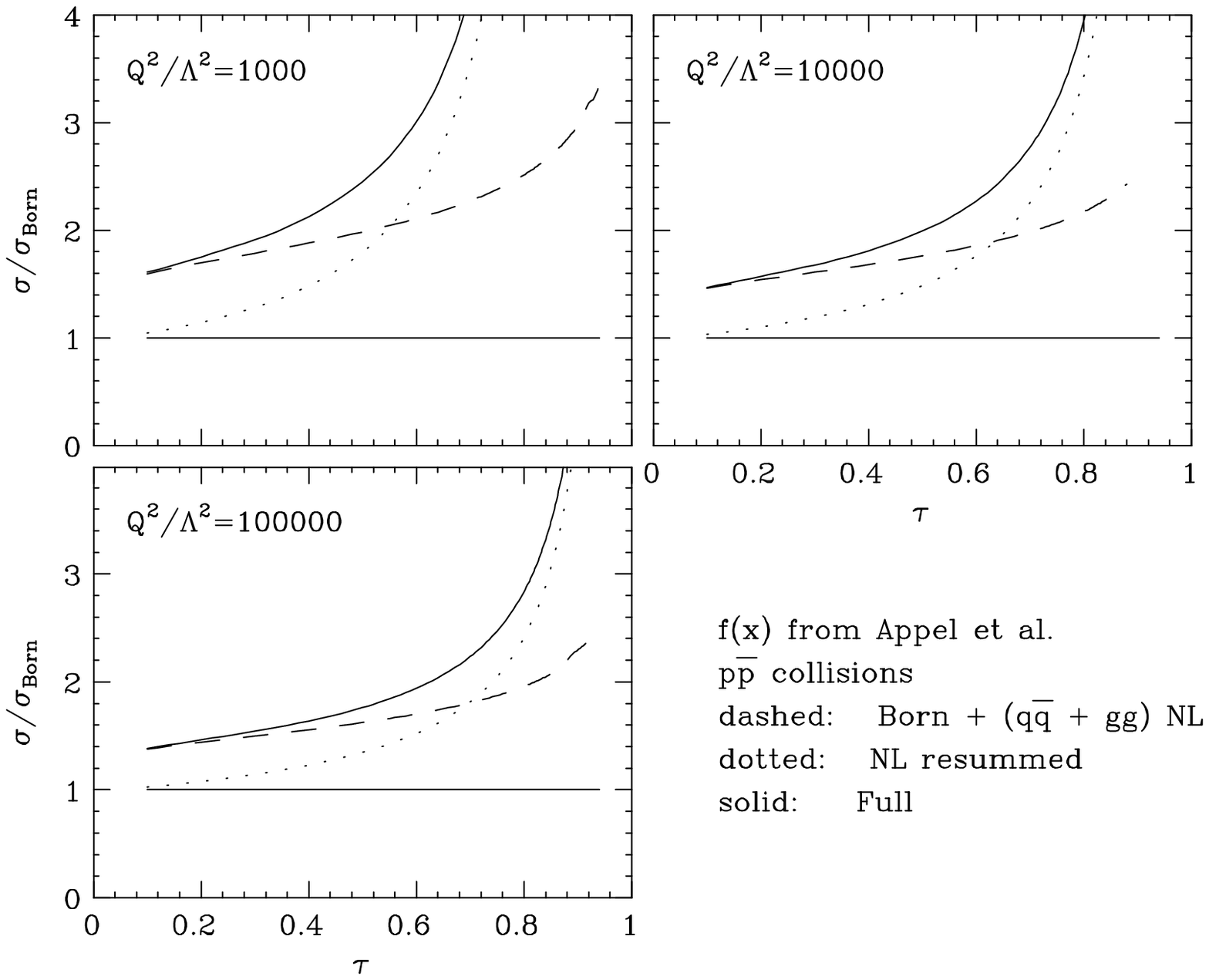,width=10cm,clip=}}
\ccaption{}{ \label{Appelppbar}
Drell--Yan pair production cross section in $p\bar{p}$
collisions, normalized to the Born result.}
\end{figure}

We will not attempt to perform a detailed phenomenological analysis
of resummation effects in Drell--Yan pair production in the present work.
We will, however, assess the effect of resummation in the particularly
simple example of the structure functions reported in Appendix B of
ref.~\cite{Appel}, for proton-antiproton collisions,
and we will compare our results with the ones
given there.
In fig.~\ref{Appelppbar} we report the cross section as a function
of $\tau=Q^2/S$, normalized to the Born cross section. With this
set of structure functions, which are $Q^2$-independent, the $K$-factor
depends only upon the ratio $Q^2/\Lambda^2$.
The dashed curve is the NL cross section,
including both the $q\bar{q}$ and $qg$
subprocesses (given in ref.~\cite{AEM}).
The dotted curve is the NL-resummed cross section,
without the inclusion of the exact ${\cal O}(\as)$ result.
The full curve is obtained by adding to the NL cross section
the NL-resummed contributions, after having subtracted the terms up
to ${\cal O}(\as)$, that is to say
\beqn
\sigma(\tau)&=&\sigma^{(q\bar{q})}_{\rm Born}(\tau)
+\frac{\as}{2\pi}\left[\sigma^{(q\bar{q})}_1(\tau)+\sigma^{(qg)}_1(\tau)\right]
\nonumber \\
&+&\left[\sigma_{\rm res}(\tau)-\sigma_{\rm res}(\tau)\Big|_{\as=0}-\as
\frac{\partial}{\partial \as}\sigma_{\rm res}(\tau)\Big|_{\as=0}\right]\;.
\eeqn
Our result is consistent with the result of ref.~\cite{Appel},
where analogous figures are given.
There is however one important difference. The cut-off method
used there to overcome the problem of the Landau singularity
introduces an extra uncertainty, which is given
by the spread of the various curves obtained using different cut-offs.
This spread decreases roughly as $1/Q$ as $Q^2$ increases.
Accounting for the fact that in ref.~\cite{Appel} also the so-called
$\pi^2$ terms are exponentiated,
our result is consistent with their band. It does not, however,
agree with the central value, which is smaller
in our case.
Furthermore, even at the lowest energy, and for the very large
value of $\tau=0.5$, the fully resummed cross section is only 10\% larger
than the next-to-leading one.

\section{Heavy flavour production}\label{hfp}
\newcommand\rhoh{\rho_{\rm\scriptscriptstyle H}}
We will follow closely the notation of ref.~\cite{NDE}.
The heavy flavour production cross section is given by the formula
\beq \label{HVQCrossSection}
\sigma=\int_0^1 dx_1\,dx_2 F(x_1) F(x_2)\;\hat{\sigma}
\left(\frac{\rhoh}{x_1\,x_2}\right)\;,\quad\rhoh=\frac{4m^2}{S}\;,
\eeq
where $m$ is the mass of the heavy quark and $S$ is the square of the
total centre-of-mass energy.
As before, for notational convenience, we have dropped here
the parton indices. The scale dependence is also not shown explicitly.
The partonic cross section depends also upon the heavy quark mass.
Here we indicate explicitly only its dependence upon $\rhoh$, which
embodies its dependence upon the partonic centre-of-mass energy
squared $s=x_1\,x_2\,S$.
In order to include the effects of soft radiation
we parallel as closely as possible
the approach followed in the Drell--Yan case. The leading logarithmic (LL)
soft corrections in heavy flavour production
have the same structure as in the Drell--Yan case \cite{NDE,Laenen},
since there are
no collinear singularities arising from final state gluon radiation.
We define
\beqn
\frac{d\sigma}{d\tau}&=&\int_0^1 dx_1\,dx_2 F(x_1) F(x_2)\;
\delta(\tau-x_1\,x_2\,)
\hat{\sigma}\left(\frac{\rhoh}{\tau}\right)\nonumber \\
&=&\hat{\sigma}\left(\frac{\rhoh}{\tau}\right)
\frac{1}{2\pi i}\int_{C-i\infty}^{C+i\infty}\; F^2_{N}\;\tau^{-N}\; dN\;.
\eeqn
The inclusion of soft effects can now be performed as in the Drell--Yan case
\beq
\frac{d\sigma^{(\rm res)}}{d\tau}
=\hat{\sigma}\left(\frac{\rhoh}{\tau}\right)
\frac{1}{2\pi i}\int_{C-i\infty}^{C+i\infty}\; F^2_{N}\;
\Delta_{N}^{HF}\;
\tau^{-N}\; dN\;.
\eeq
Using now the identity
\beq
\sigma=\int_{\rhoh}^\infty\; d\tau \;\frac{d\sigma}{d\tau}
\eeq
and defining as usual
\beq
\hat\sigma_N=\int_0^1 \frac{dz}{z}\;z^N \;\hat\sigma(z)
\eeq
we get immediately
\beq \label{MPHQ}
\sigma^{(\rm res)}=\frac{1}{2\pi i}
\int_{C-i\infty}^{C+i\infty}\; F^2_{N}\,\Delta_{N}^{HF}\;
\hat\sigma_{N-1}\; \rhoh^{-(N-1)}\; dN\;.
\eeq

The partonic cross section (after
subtraction of collinear singularities) is
\beq\label{sighq}
{\hat \sigma}_{ij}(s, m^2, \mu^2) \equiv \frac{\as^2(\mu^2)}{m^2} \;
f_{ij}(\rho, \mu^2/m^2) \;\;,
\eeq
where $\mu$ is the factorization scale (the renormalization scale is set equal
to the factorization scale) and the dimensionless variable $\rho$ is
\beq\label{rhodef}
\rho = \frac{4 m^2}{s} \;\;.
\eeq

The functions  $f_{ij}$ have the following perturbative expansion
\beq\label{fpert}
f_{ij}(\rho, \mu^2/m^2) = f_{ij}^{(0)}(\rho) + g_S^2(\mu^2) \left[
f_{ij}^{(1)}(\rho) + {\overline f}_{ij}^{(1)}(\rho)
\ln \frac{\mu^2}{m^2} \right] \;\;.
\eeq

The lowest-order terms in Eq.~(\ref{fpert}) are explicitly given by
($\beta \equiv \sqrt {1-\rho} \;$)
\beqn\label{fqqbar}
f_{q{\bar q}}^{(0)}(\rho) &=& \frac{\pi}{6} \;\frac{T_R C_F}{N_c} \;
\beta \rho \;( 2 + \rho) \;\;,
\\
\label{fgg}
f_{gg}^{(0)}(\rho) &=& \frac{\pi}{12} \;\frac{T_R}{N_c^2 -1} \;
\beta \rho  \left\{ 3 C_F \left[ ( 4 + 4 \rho - 2 \rho^2) \;
\frac{1}{\beta} \;
\ln \frac{1+\beta}{1-\beta} - 4 - 4 \rho \right] \right. \nonumber \\
&+& C_A \left. \left[ 3 \rho^2 \;\frac{1}{\beta} \;\ln \frac{1+\beta}{1-\beta}
- 4 - 5 \rho \right] \right\} \;\;,
\eeqn
and $f_{ij}^{(0)}(\rho) = 0$ for all the other parton channels.

The $N$-moments of the expressions in Eqs.~(\ref{fqqbar}) and (\ref{fgg})
are as follows
\beqn\label{fqqbarN}
f_{q{\bar q}, \,N}^{(0)} &=& \frac{\pi^{\frac{3}{2}}}{4} \;
\frac{T_R C_F}{N_c} \; \frac{\Gamma(N+1)}{\Gamma(N + 7/2)} \; (N+2) \;\;,
\\ \label{fggN}
f_{gg, \,N}^{(0)} &=& \frac{\pi^{\frac{3}{2}}}{4} \;
\frac{T_R}{N_c^2 -1} \;\frac{\Gamma(N+1)}{\Gamma(N + 5/2)} \;\frac{1}{N+3}
\nonumber \\
&\times& \left[ 2 C_F \;\frac{N^3 + 9 N^2 + 20 N + 14}{(N+1)(N+2)}
\; - C_A \;\frac{N^2 + 8 N + 11}{2 N + 5} \right] \;\;.
\eeeq
The resummation effects are embodied in the factor $\Delta_N$.
While in the Drell--Yan case this resummation factor is only associated
with the $q\bar{q}$ subprocess, in the heavy flavour case both $q\bar{q}$
and $gg$ subprocesses are involved. The explicit expression
of $\Delta_N$ will therefore depend upon the subprocess.
To {\em leading} logarithmic accuracy we have
($\as \equiv \as(m^2)$)
\beq\label{deltahf}
\ln \Delta_{ij, \,N}^{HF}(m^2) = \ln N \;g_{ij, \,1}(\bzero\as\ln N)
 + {\cal O}(\as^k \ln^{k}N)\;,
\eeq
where the functions $g_{ij, \,1}$ are related to the function $g_1$
in Eq.~(\ref{g0}) by simple colour factors. More precisely we have
\beq\label{g1hf}
g_{q{\bar q}, \,1}(\lambda) = g_1(\lambda) \;\;, \;\;\;\;
g_{gg, \,1}(\lambda) = \frac{C_A}{C_F} \;g_1(\lambda) \;\;.
\eeq
The factorization scheme dependence is of course contained in $g_1$.
We will therefore take eq.~(\ref{MPHQ}) with $C=C_{\rm MP}$
as our MP formula for heavy flavour resummed cross sections.

\section{Heavy Flavour cross section: phenomenological results}
\begin{figure}[htb]
  \centerline{\psfig{figure=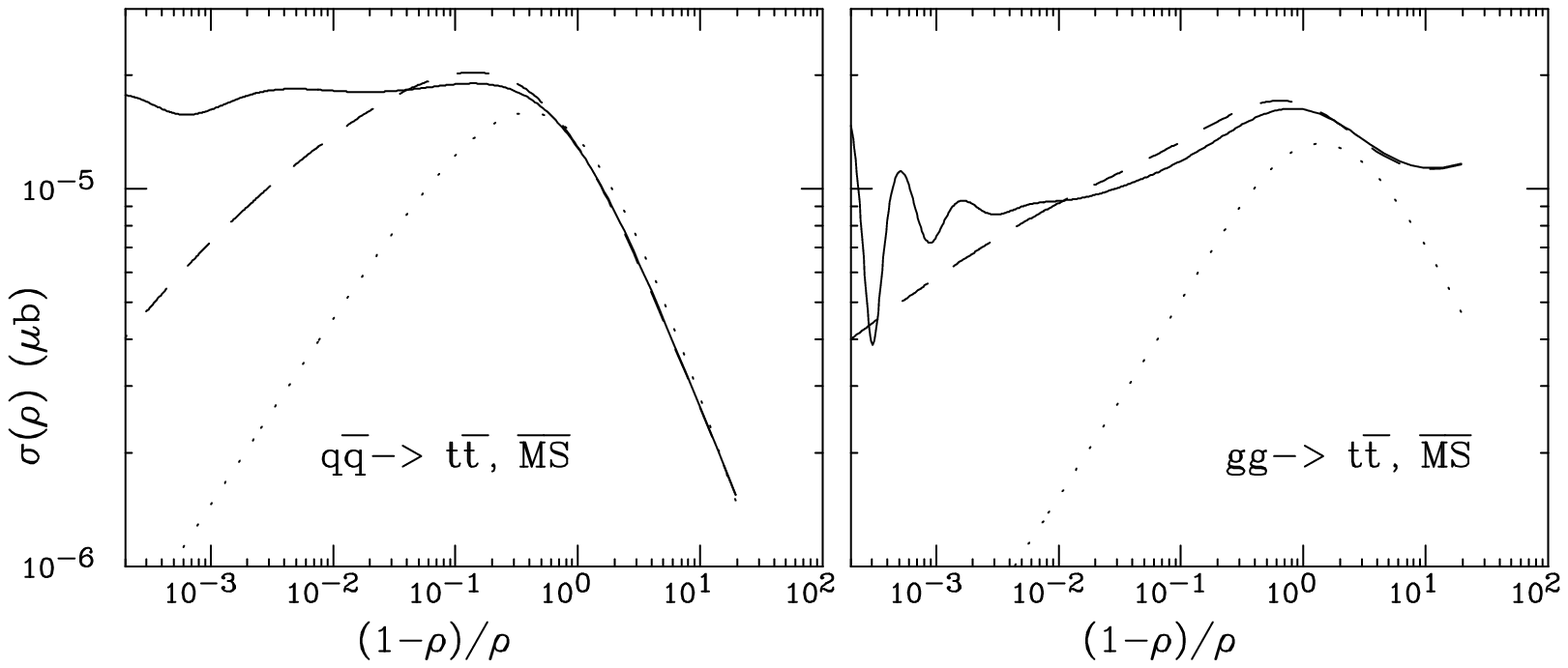,width=\textwidth,clip=}}
  \ccaption{}{
    \label{sigpart} Partonic cross section for the production of a
    175~GeV heavy quark pair.  $q\bar q$ initial state (left) and $gg$
    initial state (right). The dotted lines are the Born result, the
    dashed lines the NLO result and the solid line the resummed
    result.  \lambdamsb=152~MeV.}
\vskip 0.5cm
\centerline{\psfig{figure=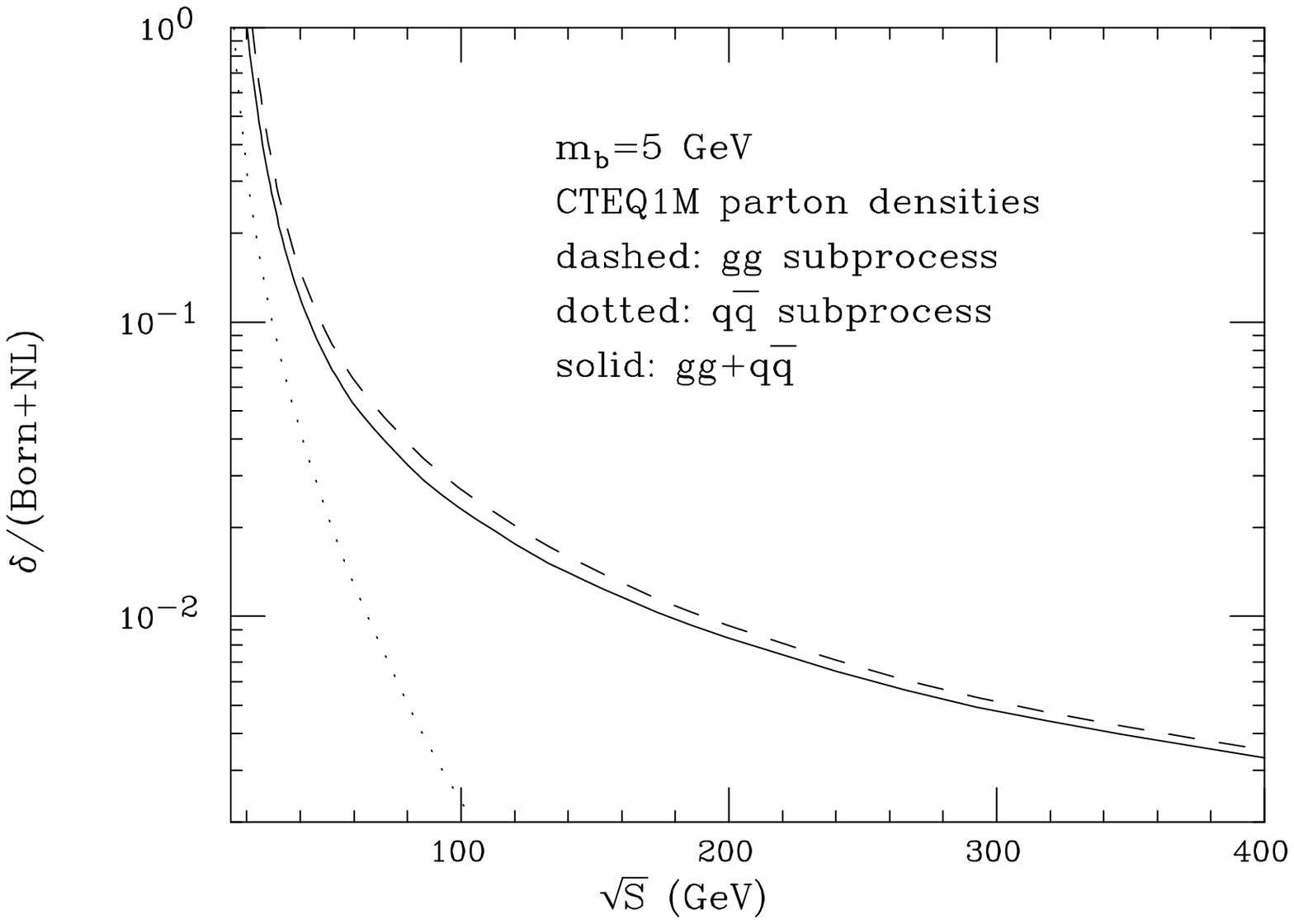,width=0.7\textwidth,clip=}}
\ccaption{}{ \label{bot-kfac}
Contribution of gluon resummation at order $\as^4$ and higher, relative to the
NLO result, for the individual channels and for the total,
for bottom production
as a function of the CM energy in $pp$ collisions.}
\end{figure}

In this section we present some phenomenological applications of the
resummation formulae presented above.  To start with, we present in
fig.~\ref{sigpart} the partonic cross sections for production of a pair of
heavy quarks of mass $m_Q=175$~GeV, plotted as a function of
$\eta=(1-\rho)/\rho$. The figures show the Born, NLO and resummed results for
both the $gg$ and $q\bar q$ initial states. The resummed partonic cross
section is defined as the sum of the contributions of order $\as^4$ and higher
from eq.~(\ref{MPHQ}) and the fixed order NLO result.  These figures can be
compared with similar ones in ref.~\cite{BergerContopanagos}. Notice that while
in that work the growth of the resummed cross section at small $\eta$ is
damped by a cut-off, in our approach the growth at small $\eta$ is
automatically controlled. Notice also that the resummed cross section is a
smooth function of $\rho$ down to values of $1-\rho$ of the order of
$10^{-3}$, which is of the order of the ratio $\Lambda/m_Q$.  Since the energy
of the soft radiation is of the order of $(1-\rho)m_Q$, at these values
of $\rho$ it is numerically of the order of $\Lambda$.  This is the point
at which the Landau pole is expected to influence the results, and
non-perturbative physics to set on.  We interpret this behaviour as a
confirmation of the correctness of our procedure. It would make no sense to
cut off the partonic cross section at values of $\rho$ corresponding to
soft gluon radiation of several GeV.
For $1-\rho<10^{-3}$ the resummed
cross section starts to oscillate, but it remains integrable.


\begin{figure}
\centerline{\epsfig{figure=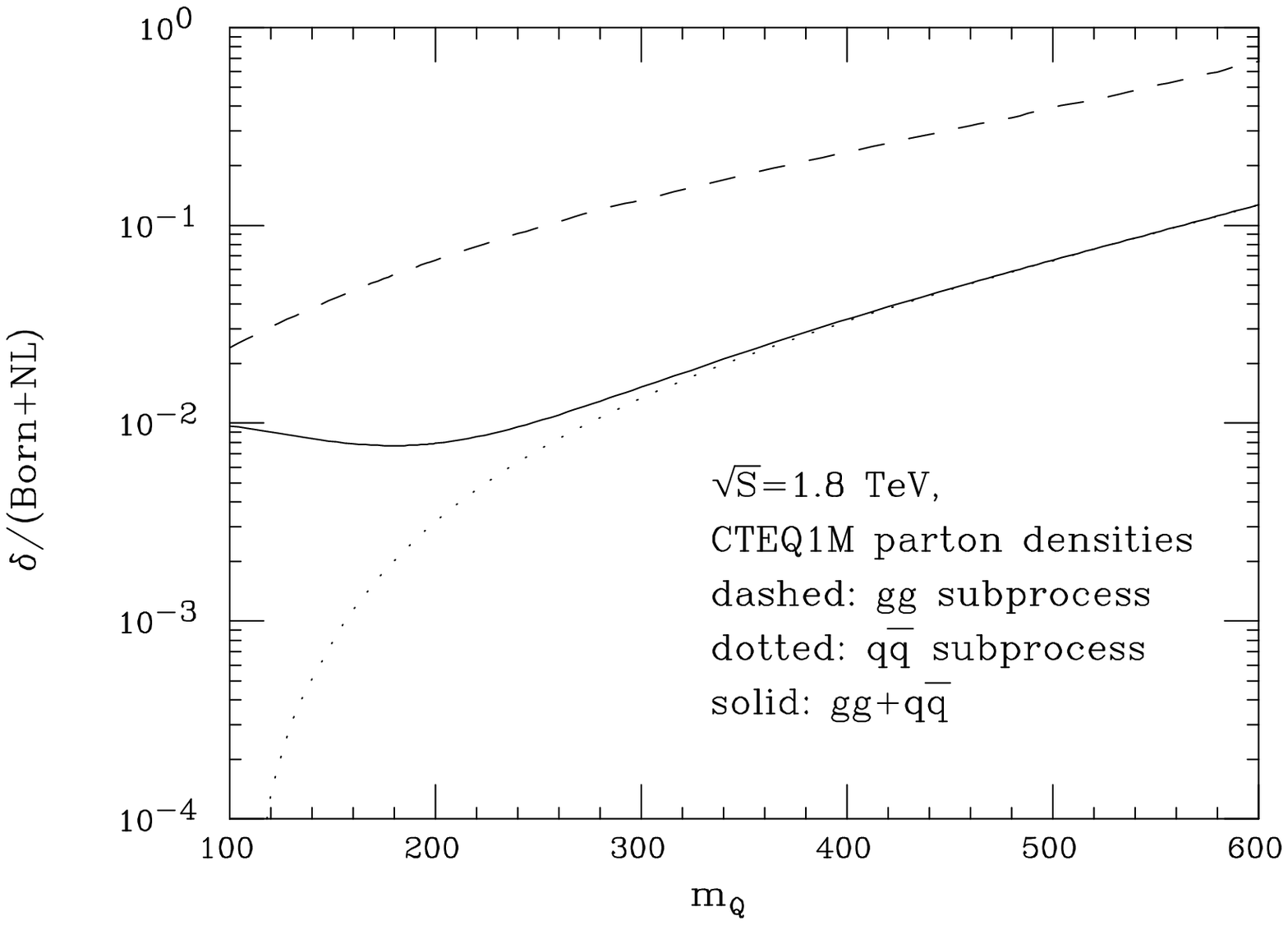,width=0.7\textwidth,clip=}}
\ccaption{}{ \label{frtev}
Contribution of gluon resummation at order $\as^4$ and higher, relative to the
NLO result, for the individual subprocesses and for the total,
as a function of the top mass in $p\bar p$ collisions at 1.8 TeV. }
\end{figure}
\begin{figure}
\centerline{\epsfig{figure=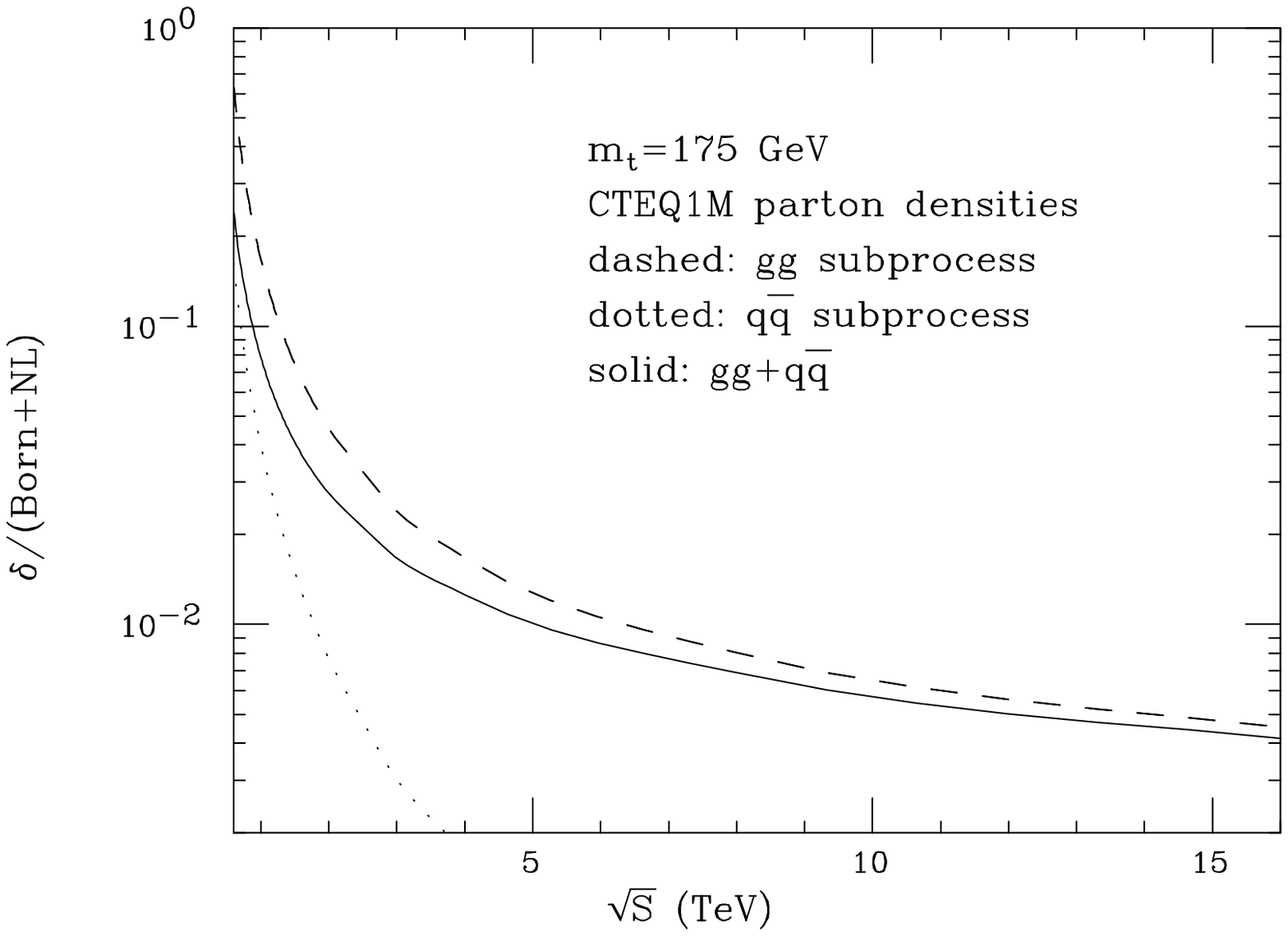,width=0.7\textwidth,clip=}}
\ccaption{}{ \label{frlhc}
Contribution of gluon resummation at order $\as^4$ and higher, relative to the
NLO result, for the individual channels and for the total,
as a function of the CM energy in $pp$ collisions. }
\end{figure}

In all our phenomenological studies of resummation effects in heavy
flavour production we have used the structure function set CTEQ1M
\cite{CTEQ1}.  The importance of the resummation effects is
illustrated in figs.~\ref{bot-kfac}, \ref{frtev} and \ref{frlhc},
where we plot the quantities
\beq \label{deltadef} \frac{\delta_{\rm
    gg}}{\sigma^{(gg)}_{\rm NLO}}\,,\quad \frac{\delta_{\rm
    q\bar{q}}}{\sigma^{(q\bar{q})}_{\rm NLO}}\,,\quad
\frac{\delta_{\rm gg}+\delta_{\rm q\bar{q}}}{\sigma^{(gg)}_{\rm
    NLO}+\sigma^{(q\bar{q})}_{\rm NLO}}\;.
\eeq
Here $\delta$ is equal
to our MP resummed hadronic cross section in which the terms of order
$\as^2$ and $\as^3$ have been subtracted, and $\sigma_{(\rm NLO)}$ is
the full hadronic NLO cross section.
The results for $b$ at the Tevatron can be easily inferred
from fig.~\ref{bot-kfac}, since the $q\bar{q}$ component is negligible
at Tevatron energies.

For top production, we see that in most configurations of practical
interest, the contribution of resummation is very small, being of the order
of 1\% at the Tevatron.
A complete review of top quark production at the Tevatron,
based upon these findings, has already been given in
ref.~\cite{cmnt21}. We also observe that, for top production at the LHC,
soft gluon resummation effects are negligible. Of course, in this last case,
there are other corrections, not included here, that may need to be
considered. Typically, since the values of $x$
involved are small in this configuration, one may have to worry
about the resummation of small-$x$ logarithmic effects \cite{smallx}.

We see from the figures that in most experimental configurations of
interest these effects are fully negligible. One noticeable
exception is $b$ production at HERAb, at $\sqrt{S}=39.2$, where
we find a 12\% increase in the cross section.
%
%
%
This correction is however well below the uncertainty due to higher order
radiative effects. For example, from the NLO calculation
with the MRSA$^\prime$ \cite{MRSAp} parton
densities and $m_b=4.75\,$GeV, we get
$\sigma_{b\bar{b}}=10.45{ +8.24  \atop -4.65}\,$nb,
a range obtained by varying the renormalization
and factorization scales from $m_b/2$ to $2m_b$. Thus
the upper band is 80\% higher than the central value, to be compared with
a 10\% increase from the
resummation effects. This result is much less dramatic than
the results of ref.~\cite{KidonakisSmith95}.

\begin{figure}
\centerline{\epsfig{figure=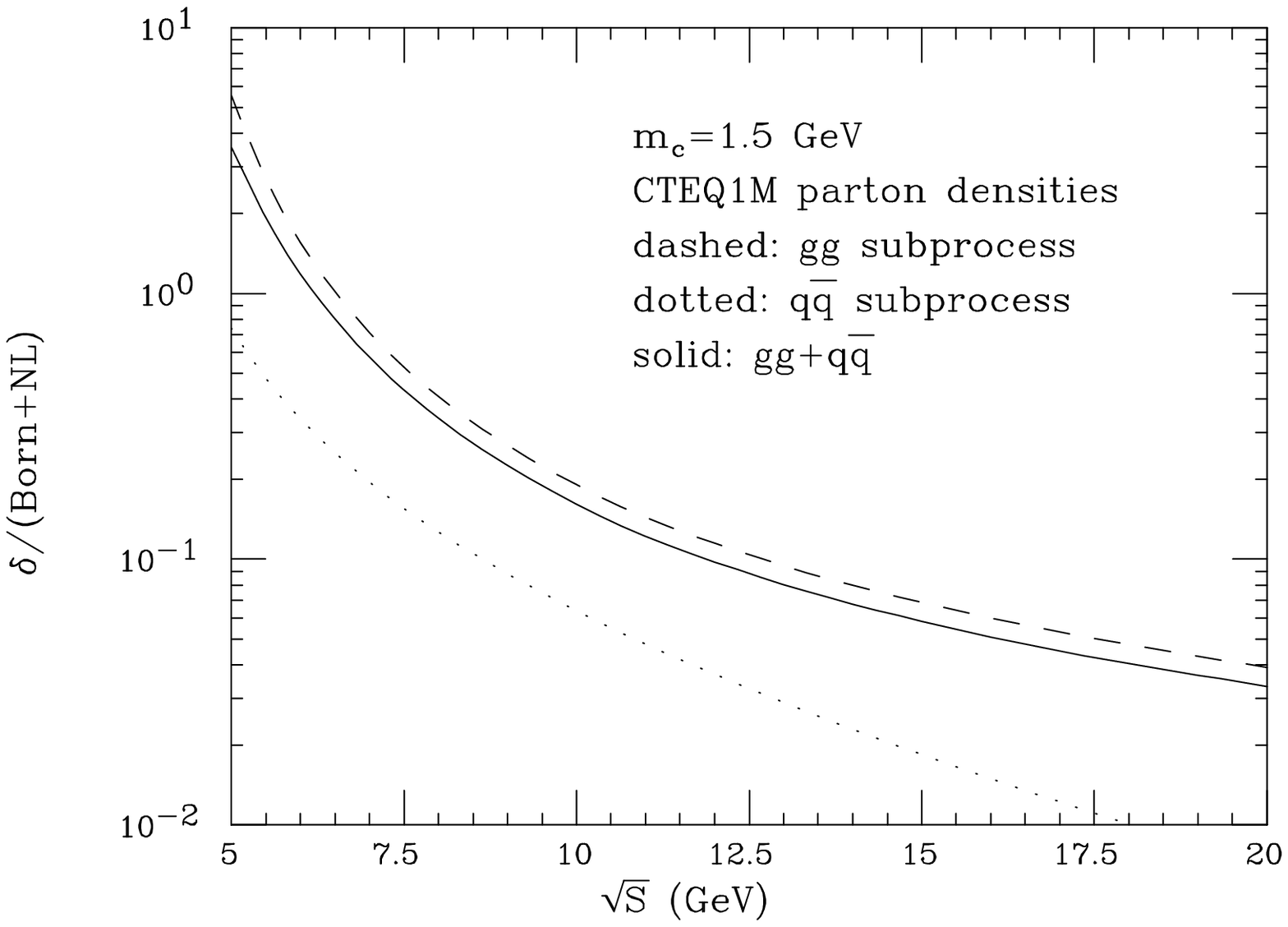,width=0.7\textwidth,clip=}}
\ccaption{}{ \label{charm-kfac}
Contribution of gluon resummation at order $\as^4$ and higher, relative to the
NLO result, for the individual channels and for the total,
for charm production
as a function of the CM energy in $pp$ collisions.}
\end{figure}

In fig.~\ref{charm-kfac} we also show the effect of resummation
in charm production. Due to the large uncertainties that plague
charm production \cite{charm},
this plot should only be considered for orientation.

We conclude with a few remarks and checks about our result.
We will focus on top production by $q\bar{q}$ annihilation,
with $m_t=175$ GeV at the
Tevatron. First of all, we checked that the full resummation
formula that we use is well approximated by its expansion
in powers of $\as$. We define
\beq
\sigma^{\rm (res)}_M=\sum_{k=2}^M \sigma_k \;,
\eeq
where $\sigma_k$ is the contribution of order $\as^k$.
We computed each term in the expansion up to order $\as^6$.
The results are displayed in table~\ref{orderbyorder}.
We see that the convergence properties of the expansion are
extremely good, and, up to the order we have probed, there is
no sign that we are near the breakdown of the expansion.
As a second observation, we notice that the term $\sigma_3$
in our resummed formula is about 7\% of the Born result.
The full ${\cal O}(\as^3)$ correction is instead 20\% of the Born term.
It is easy to illustrate the contribution of the $\sigma_3$ term
to the partonic cross section.
We start with the full NLO cross section written in the
form of eqs.~(\ref{sighq},\ref{fpert})
\beq
\hat\sigma^{(NLO)}_3(\rho)=\frac{\as}{m^2} \left( f^{(0)}_{q\bar{q}}(\rho)
+4\pi\as\, h_{q\bar{q}}(\rho) \right)
\eeq
where
\beq
h_{q\bar{q}}(\rho)=f^{(1)}_{q\bar{q}}(\rho)
+\bar{f}^{(1)}_{q\bar{q}}(\rho)\log\frac{\mu^2}{m_t^2} \; .
\eeq
The functions $f^{(0)}_{q\bar{q}}$, $f^{(1)}_{q\bar{q}}$ and
$\bar{f}^{(1)}_{q\bar{q}}$ are defined in ref.~\cite{NDE}, and $\mu$ is
the factorization and renormalization scale.
It is easy to show that the truncated resummed result
at order ${\cal O}(\as^3)$ can be obtained by using the following
partonic cross section
\beq
\hat\sigma^{(res)}_3(\rho)=\frac{\as}{m^2} \left( f^{(0)}_{q\bar{q}}(\rho)
+4\pi\as\, h^\prime_{q\bar{q}}(\rho) \right)
\eeq
where the function $h^\prime_{q\bar{q}}$ can be obtained by expanding
the resummation function $\Delta_{q\bar q,N}^{HF}(m^2)$, as given in
eq.~(\ref{deltahf}), up to order $\as$. After some simple algebra one obtains
\beq\label{h1}
 h^\prime_{q\bar{q}}(\rho) =
 \frac{C_F}{\pi^2}\int_\rho^1 dy\;f^{(0)}_{q\bar{q}}(\rho/y)
 \left[\frac{1}{\log 1/y}\left(\log\log 1/y +\gamma_E\right)\right]_+
 \; .
\eeq
The meaning of the plus distribution is as usual
\beq
\int_0^1 dy \left[G(y)\right]_+\;F(y)=\int_0^1 dy\; G(y)\;
\left(F(y)-F(1)\right)\;.
\eeq
Notice the presence of the subleading term proportional to $\gamma_E$
in the equation. This term is cancelled if we use, when performing the $x$
integration in eq.~(\ref{deltan}), the relation
$1-x^N \; = \; \theta(1-x-\frac{e^{-\gamma_E}}{N})$ (which is accurate
up to NLL~\cite{CataniTrentadue}) instead of the approximate one,
$ 1-x^N \; = \; \theta(1-x-\frac{1}{N})$.
This amounts to the substitution:
\beq
    \ln \; \Delta_{q\bar q,N}^{HF}(m^2)
  \rightarrow
    \left( 1+ \gamma_E \frac{\partial}{\partial \ln N} \right)
     \ln \; \Delta_{q\bar q,N}^{HF}(m^2) \; .
\eeq
Equation~(\ref{h1}) then becomes
\beq\label{h1p}
 h^{\prime\prime}_{q\bar{q}}(\rho) =
 \frac{C_F}{\pi^2}\int_\rho^1 dy\; f^{(0)}_{q\bar{q}}(\rho/y)
 \left[\frac{1}{\log 1/y} \log\log 1/y \right]_+\;\;.
\eeq

\begin{figure}
\centerline{\epsfig{figure=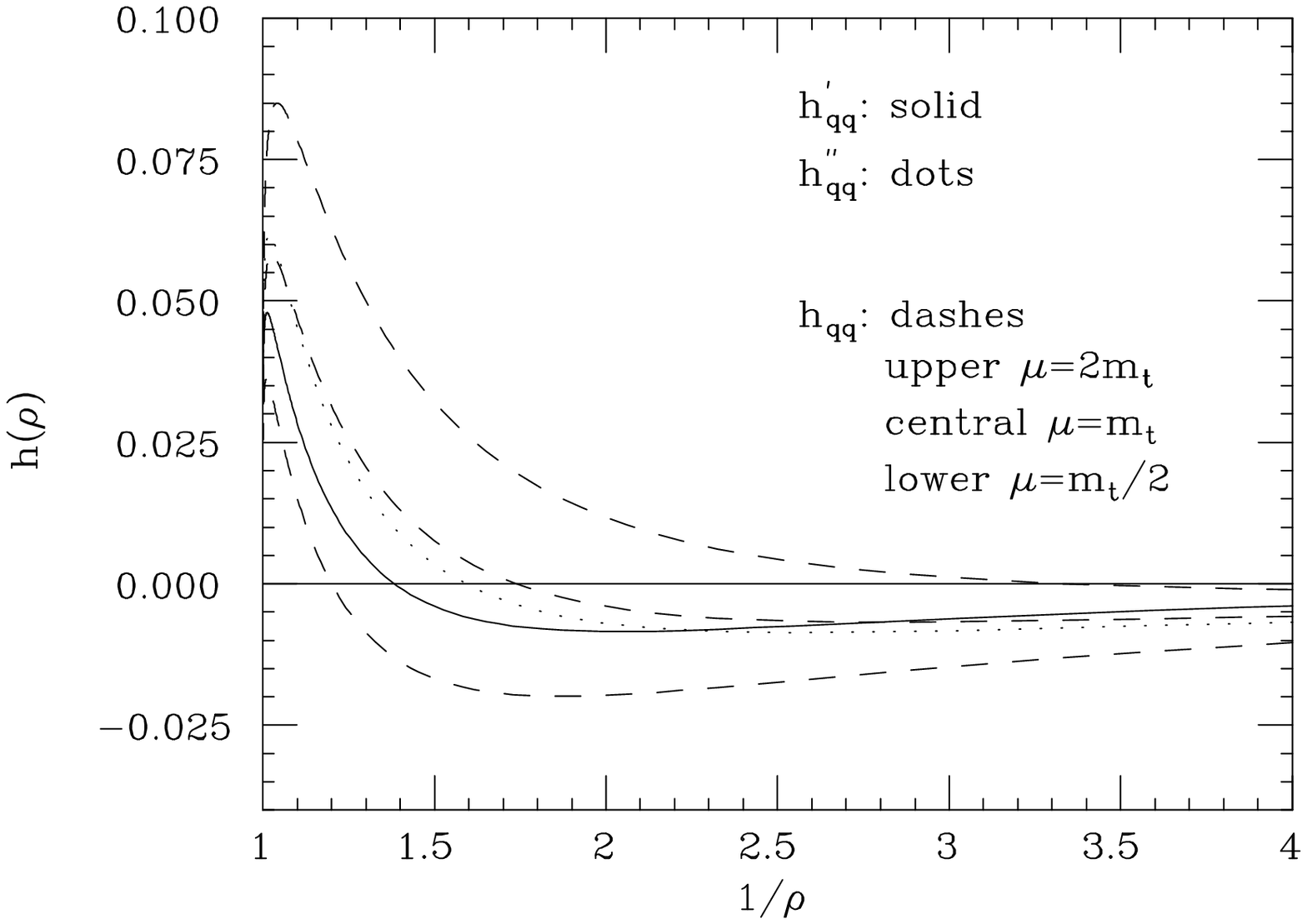,width=0.7\textwidth,clip=}}
\ccaption{}{ \label{nlsoft}
Comparison of the partonic cross section at ${\cal O}(\as^3)$
for the MP resummation formula (solid line), for the resummation formula
with the term proportional to $\gamma_E$ included (dotted line),
and for the exact ${\cal O}(\as^3)$ formula of ref.~\protect{\cite{NDE}}
with  $\mu=m_t/2$, $m_t$ and $2m_t$ (dashed lines).}
\end{figure}

In fig.~\ref{nlsoft} we give the contribution of the
$h^{(1)}_{q\bar{q}}(\rho)$ compared with the quantity that corresponds
to the full NLO correction.

  We plot the full NLO
correction for $\mu=m_t/2$, $m_t$ and $2m_t$.
\begin{table}
\begin{center}
\begin{tabular}{|l|c|c|c|c|c|c|} \hline
$k,M=$& 2 & 3 & 4 & 5 & 6 \\
\hline
$\sigma_k\;$(pb) & 3.71 & 0.256 & $8.27\times 10^{-3}$
& $-3.24\times 10^{-4}$ & $-5.88 \times 10^{-5}$ \\ \hline
$1-\sigma^{\rm (res)}_M/\sigma^{\rm (res)}$
 & 0.066 & $2.0 \times 10^{-3}$ & $- 9.63 \times 10^{-5}$ &
  $-1.48 \times 10^{-5}$ & $4.8 \times 10^{-8}$
\\ \hline
\end{tabular}
\ccaption{}{\label{orderbyorder}
Top pair production at the Tevatron via the $q\bar q$ channel, for
$m_t=175$~GeV and CTEQ1M parton densities.
Order-by-order contributions to the fully resummed formula (first
line), and accuracy of the truncated perturbative expansion relative
to the fully resummed result (second line). }
\end{center}
\end{table}
We see that the $h^\prime_{q\bar{q}}$ term is consistent with the exact
next-to-leading result, given the spread of the latter induced by the
renormalization scale dependence.
According to the choice of subleading terms,
the $\sigma_3$ term can substantially change, and can be brought
to almost coincide with the exact result for $\mu=m_t$.
In our case, for example, inclusion of the term proportional to
$\gamma_E$ in formula (\ref{g1ms}) would cancel
exactly the corresponding term appearing in formula (\ref{h1}),
which then reduces to Eq.~(\ref{h1p}). This last formula gives
a result that is very close
to the exact one for $\mu=m_t$ in the most important kinematic region
of $\rho\approx 1$,
as can be seen from fig.~\ref{nlsoft}.
This is in agreement with the findings of ref.~\cite{Laenen}, where
it was pointed out that the LL truncation of the ${\cal O}(\as^3)$
terms provide a very good approximation to the full NLO cross section,
as long as $\mu=m_t$. However it should be stressed once more that this
result is accidental, as it would not hold for a different choice of
renormalization scale.

Notice that altering the structure of the subleading terms in the
exponent of the Mellin-space coefficient function will not affect the
asymptotic properties of its perturbative expansion and the
integrability of its $x$-space MP transform.
We explored the numerical impact on the contributions of order $\as^4$ and
higher of including the
subleading terms proportional to $\gamma_E$ in the exponent of the
coefficient function.
In the notation of Eq.~(\ref{deltadef})
we get $\delta_{q\bar{q}}/\sigma_{\rm NL}=0.013$, an effect which is
larger than the one found previously, but still negligible.
The difference from the result of fig.~\ref{frtev} should  be taken as an
estimate of the uncertainty coming from the unknown next-to-leading
logarithmic terms in the exponentiated coefficient function. As such,
it is a purely perturbative uncertainty, which cannot be separated
from the uncertainty due to the change in renormalization scale or
factorization scheme. The size of these latter uncertainties,
estimated to be of the order of 10\% \cite{cmnt21}, can consistently
accommodate for the 1\% effect we found.


\section{Jet Cross Sections}\label{jcs}
In this section we present, as an additional application of our formula for
the soft gluon resummation, a study of corrections to the invariant mass
distribution of jet pairs produced in $p \bar p$ collisions at 1.8 TeV.  The
interest in the effects of resummation on the behaviour of jet cross sections
at large energy is prompted by the discrepancy between the single-inclusive
jet-\pt\ distribution at large \pt, as measured by CDF \cite{cdfjet}, and the
result of the NLO QCD predictions \cite{nlojet}.  For simplicity we will study
the effects of soft gluon resummation on the invariant mass distribution of
the jet pair, which is, from a theoretical point of view, very close to the
Drell--Yan pair production.  Observe that other distributions,
such as the $\pt$
of the jet, have a rather different structure from the point of view of soft
gluon resummation. In fact, while the jet pair mass is only affected by the
energy degradation due to initial state radiation, the $\pt$ of the jet may
also be affected by the transverse momentum generated by initial state
radiation, and by the broadening of the jet due to final state radiation.

A study of the jet pair mass distribution is not of purely academic interest,
since also for this variable an analogous discrepancy between data
and theory has been observed \cite{cdfmass}.
Studies of resummation effects in the inclusive $\pt$ spectrum of jets
are in progress (M. Greco and P. Chiappetta, private communication).

Before presenting the results, we
briefly discuss the key elements of the calculation.
Contrary to the case of Drell--Yan and heavy flavour production, light jets are
produced at Born level from all possible initial states, $gg$, $qg$, $qq$
and $q\bar q$.
The resummed invariant mass distribution of dijets, at LL order, is
therefore given by
\beq\label{sjetres}
\frac{d\sigma^{(\rm res)}}{d\tau}
=\sum_{i,j \in \{q,\bar q,g\}}
\hat{\sigma}_{ij}\left(\hat{s},\theta^*_{min}\right)
\frac{1}{2\pi i}\int_{C-i\infty}^{C+i\infty}\; F^{i,N}F^{j,N}\;
\Delta_{ij,N}^J\;
\tau^{-N}\; dN\;,
\eeq
where $\hat{s}=S\tau= m^2_{JJ}$.
The functions $\Delta_{ij,N}^J$ are given by eqs.~(\ref{deltahf}) and
(\ref{g1hf}), supplemented by
\beeq{g1qg}
\ln \Delta_{qg, \,N}^J(m^2_{JJ}) &=& \ln N \;g_{qg, \,1}(\bzero\as\ln N)
 + {\cal O}(\as^k \ln^{k}N)
\\
g_{qg, \,1}(\lambda) &=& \frac{C_A+C_F}{2C_F} g_1(\lambda)\;.
\eeeq
The partonic Born cross section $\hat{\sigma}_{ij}\left(\hat{s},
\theta^*_{min}\right)$ depends on the range of integration for the partonic
centre-of-mass  scattering angle $\theta^*$ ($\theta^*$ is related to the
rapidity difference $\eta^*=(\eta_1 -\eta_2)/2$ of the two jets by
$\sin\theta^* =1/\cosh \eta^*$). To avoid the Rutherford
singularity, we will keep this angle strictly larger than zero. While the
absolute production rate depends on the choice of $\theta^*_{min}$, we will now
show that the $K$-factor, \ie\ the ratio of the resummed cross section to the
LO one is to good approximation independent of it. Therefore the results for
the $K$-factor will be rather independent of the details of the experimental
cuts.
It is a well-known fact \cite{Maxwell} that the LO amplitudes for parton-parton
scattering processes are related to one another, in the small scattering angle
limit, as follows
\beq \label{ESFapp}
     d\hat{\sigma}_{gg} \; : \;  d\hat{\sigma}_{qg} \; : \; d\hat{\sigma}_{qq}
     \; = \;
     1 \; : \; \frac{4}{9} \; : \; \left ( \frac{4}{9} \right )^2 \; ,
\eeq
where the indices refer to the pair of partons in the initial state, $q$ being
an arbitrary quark (or antiquark) flavour. Even at 90$^\circ$ this
approximation is good to about 10\%, becoming better and better when the
cross sections are integrated over a larger and larger range of the
scattering angle.
Using eq.~(\ref{ESFapp}), we can rewrite the expression for the resummed jet
invariant mass cross section as follows
\beq
\frac{d\sigma^{(\rm res)}}{d\tau}
=
\hat{\sigma}_{gg}\left(\hat{s},\theta^*_{min}\right)
\frac{1}{2\pi i}\int_{C-i\infty}^{C+i\infty}\; H^2_{N}
\tau^{-N}\; dN\;,
\eeq
with
\beq
H_{N}=F_{g,N} N^{\frac{C_A}{2C_F}g_1(\lambda)} +
    \frac{4}{9}
    \sum_{i\in\{q_f,\bar{q}_f\}} F_{i,N} N^{\frac{1}{2}g_1(\lambda)}
    \; .
\eeq
The $K$-factor is therefore given by the following expression, in which all
dependence upon the acceptance cut $\theta^*$ has disappeared
\beq
\dispfrac{\frac{d\sigma^{(\rm res)}}{d\tau}}{\frac{d\sigma^{(\rm LO)}}{d\tau}}
=
\dispfrac{\frac{1}{2\pi i}\int_{C-i\infty}^{C+i\infty}\; H^2_{N}
\tau^{-N}\; dN}
     {\frac{1}{2\pi i}\int_{C-i\infty}^{C+i\infty}\; E^2_{N}
\tau^{-N}\; dN}\;.
\eeq
$E_{N}$ is the Mellin transform of the standard {\it effective structure
function}, defined by
\beq
E_{N}=F_{g,N} +
    \frac{4}{9}
    \sum_{i\in\{q_f,\bar{q}_f\}} F_{i,N}
    \; .
\eeq
While a more accurate implementation of the experimental acceptance will
eventually be necessary\footnote{For instance, at small $\theta^*$ a more
refined resummation procedure would lead to the replacement $M_{jj}^2 \to
M_{jj}^2\sin^2(\theta^*/2)$ in the hard scale of the resummed coefficient
functions $\Delta^J_{ij,N}$ in eqs.~(\ref{sjetres},\ref{g1qg}).},
our approximation is more than adequate to provide a first estimate of the
resummation effects.

\begin{figure}
\centerline{\epsfig{figure=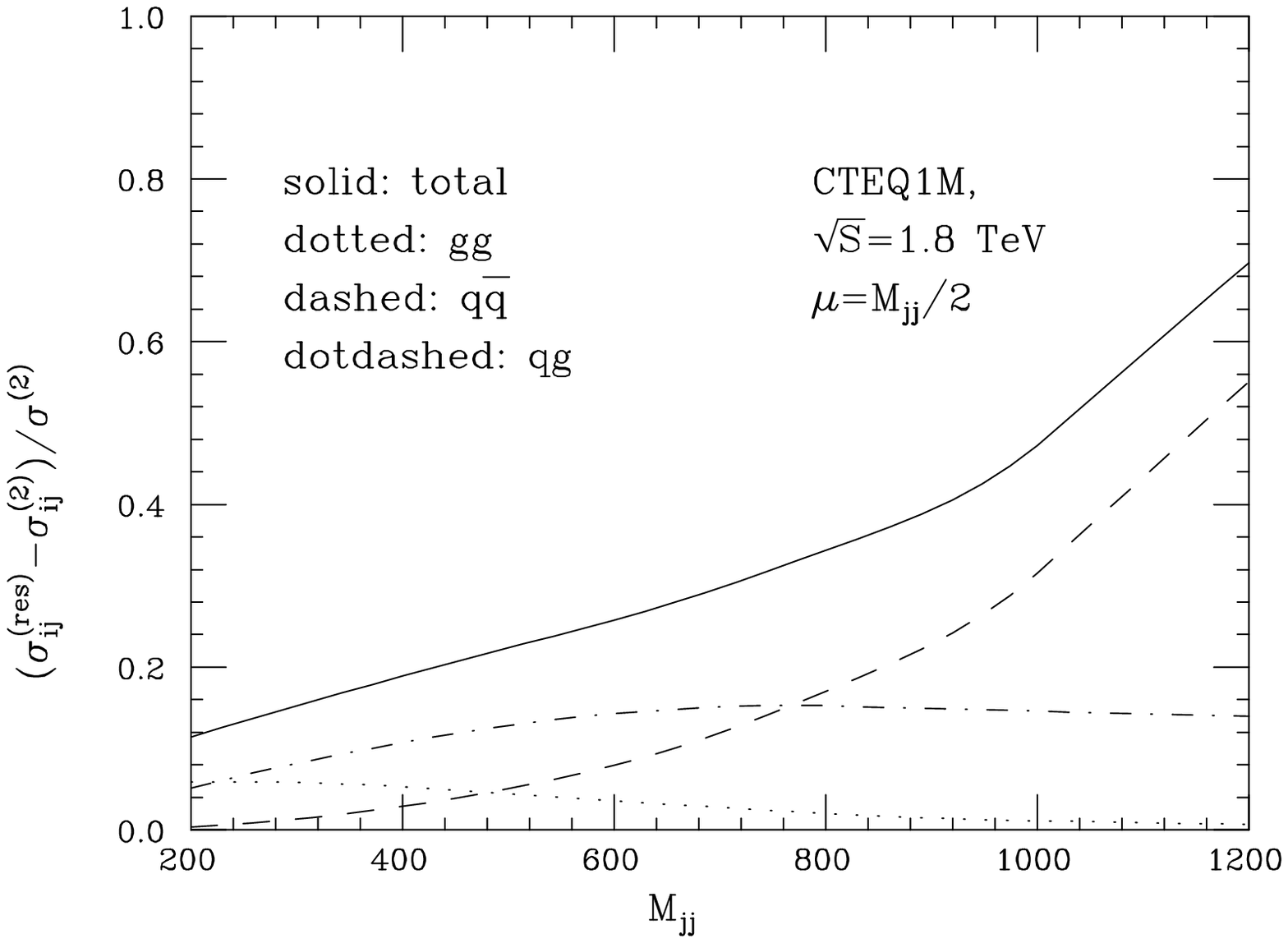,width=0.66\textwidth,clip=}}
\ccaption{}{ \label{jetcteq1}
Contribution of gluon resummation at order $\as^3$ and higher, relative to the
LO result, for the invariant mass distribution of jet pairs at the Tevatron.}
\vskip 1cm
\centerline{\epsfig{figure=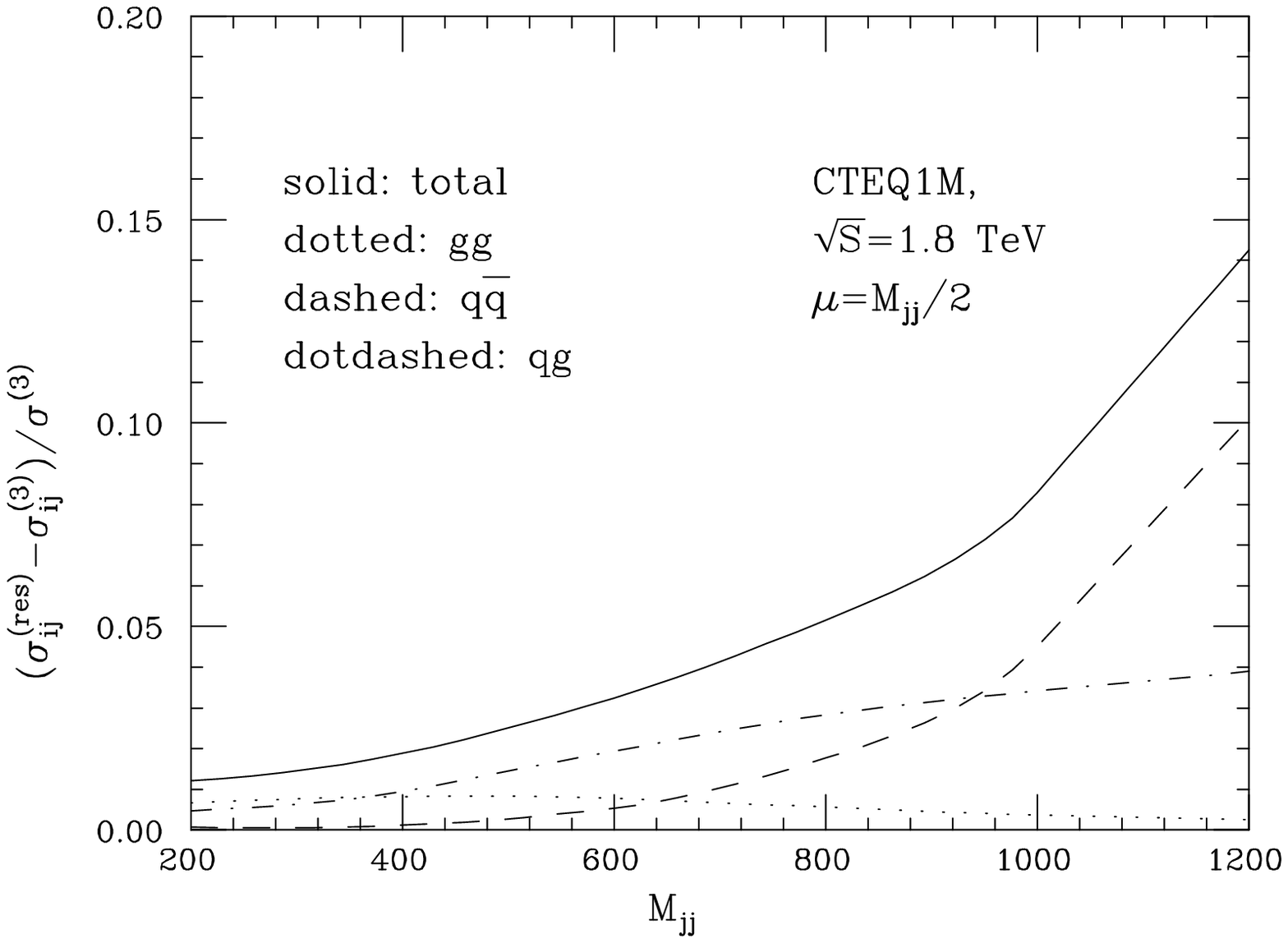,width=0.66\textwidth,clip=}}
\ccaption{}{ \label{jetcteq2}
Contribution of gluon resummation at order $\as^4$ and higher, relative to the
truncated ${\cal O}(\as^3)$ result,
for the invariant mass distribution of jet pairs at the Tevatron.}
\end{figure}

In fig.~\ref{jetcteq1} we show the following quantities
\beq
\frac{\delta^{(3)}_{\rm gg}}{\sigma^{(2)}}\,,\quad
\frac{\delta^{(3)}_{\rm qg}}{\sigma^{(2)}}\,,\quad
\frac{\delta^{(3)}_{\rm q\bar{q}}}{\sigma^{(2)}}\,,\quad
\frac{\delta^{(3)}_{\rm gg}+\delta^{(3)}_{\rm qg}+\delta^{(3)}_{\rm
q\bar{q}}}{\sigma^{(2)}}
\eeq
where $\delta^{(3)}$ is equal to
our MP resummed hadronic cross section
in which the terms of order $\as^2$
have been subtracted, and
$\sigma^{(2)}=\sigma^{(2)}_{\rm gg}+\sigma^{(2)}_{\rm qg}
 +\sigma^{(2)}_{\rm q\bar{q}}$
is the full hadronic LO cross section
(of order $\as^2$).
We use as a reference renormalization and factorization
scale for our results $\mu=M_{jj}/2$. Notice that for
large invariant masses the effects of higher orders are large.
To understand how much is due to the first order corrections
(which are exactly calculable \cite{nlo2jet})
and how much is due to corrections of order
$\as^4$ and higher, we show the following quantities in fig.~\ref{jetcteq2}
\beq
\frac{\delta^{(4)}_{\rm gg}}{\sigma^{(3)}}\,,\quad
\frac{\delta^{(4)}_{\rm qg}}{\sigma^{(3)}}\,,\quad \frac{\delta^{(4)}_{\rm
    q\bar{q}}}{\sigma^{(3)}}\,,\quad \frac{\delta^{(4)}_{\rm
    gg}+\delta^{(4)}_{\rm qg}+\delta^{(4)}_{\rm q\bar{q}}}{\sigma^{(3)}} \; ,
\eeq
where $\delta^{(4)}$ is now equal to the MP resummed hadronic cross section
with terms of order $\as^3$ subtracted, and $\sigma^{(3)}$ is an approximation
to the full NL cross section, summed over all subprocesses, obtained by
truncating the resummation formula at order $\as^3$.  This figure shows that
indeed most of the large $K$ factor is due to the pure NLO corrections, with
the resummation of higher-order soft gluon effects contributing only an
additional 10\% at dijet masses of the order of 1~TeV.  Removing
the subleading term proportional to $\gamma_E$ from the exponent of
the coefficient function, in the spirit of the discussion at the end
of the previous section, we found only a slight increase of the
contribution from the terms of order $\as^4$ and higher (an
additional absolute 5\% for dijet masses of 1~TeV).

As we will discuss in more detail in the next section, these results
should only be taken as an indication of the order of magnitude of the
correction, since we have not included here a study of the resummation
effects on the determination of the parton densities.  From this
preliminary study it seems however unlikely that the full 30--50\%
excess reported by CDF for jet \pt's in the range 300--450~GeV could
be explained by resummation effects in the hard process. It is
possible that the remaining excess is due to the poor knowledge of the
gluon parton densities at large $x$, an idea pursued by the CTEQ group
\cite{tung}. This problem has been studied also in ref.~\cite{GMRS}.

\section{A few additional remarks}\label{remark}
A fully
consistent treatment of the resummation effects requires the use of parton
densities that
\begin{itemize}
\item[$(i)$]
are extracted from low-energy data by taking into account
the resummation effects for the corresponding scattering process;
\item[$(ii)$]
are evolved in $Q^2$ using resummed anomalous dimensions.
\end{itemize}
In both
steps $(i)$ and $(ii)$, the Minimal Prescription should be implemented.

In the case, for example, of jet production, it is quite possible that
resummation effects significantly influence the determination of the large-$x$
structure functions from low energy data. The Tevatron jets of the highest
energy probe the partonic densities at values of $x$ of the order of 0.5,
which one could argue are far enough from the $x\to 1$ region to make the
Sudakov effects small.  However, the large values of $Q^2$ involved (of the
order of $10^6 \;{\rm{GeV}}^2$) are such that the $Q^2$ evolution of the
parton densities from low energy is significant, and we are therefore
sensitive to the input structure functions measured at low $Q^2$ in regions of
$x$ significantly larger than 0.5. It would therefore be important to
reexamine the extraction of the large-$x$ non-singlet structure functions, in
the light of the resummation results for the DIS process, before firmer
conclusions can be drawn on the significance of the present jet cross section
discrepancy.

In the $Q^2$ evolution of the parton densities, one has to use the equation
\beq\label{eveq}
\frac {d F_{i,N}(Q^2)}{d \ln Q^2} = \sum_j \gamma_{ij,N}(\as(Q^2))
\;F_{j,N}(Q^2)
\;\;,
\eeq
with anomalous dimensions $\gamma_{ij,N}$ that include the  resummation of
logarithmic contributions to the same accuracy as in the coefficient function.
For instance, the analogue of Eq.~(\ref{lnd}) is the following expansion
\beq\label{adexp}
\gamma_{ij,N}(\as) = \gamma_{ij}^{(1)}(b_0 \as \ln N) +
\as \;\gamma_{ij}^{(2)}(b_0 \as \ln N) + \dots \;\;.
\eeq

The leading and next-to-leading resummed anomalous dimensions
$\gamma_{ij}^{(1)}$ and $\gamma_{ij}^{(2)}$ have, in general,
singularities related to the Landau pole
and these singularities (like those in Eqs.~(\ref{g0}), (\ref{g1}), and
(\ref{g0ms}), (\ref{g1ms})) depend on the factorization scheme in which the
parton
densities are defined. Thus, for the purpose of implementing the Minimal
Prescription, it is particularly convenient to use the ${\overline {\rm MS}}$
definition of the parton densities. Indeed, in the  ${\overline {\rm MS}}$
factorization scheme the resummed anomalous dimensions have no singularity
associated with the Landau pole and, more precisely, they have the following
simple form \cite{CataniTrentadue,Korchemsky}
\beeq{adms}
\gamma_{qq,N}^{\overline {\rm MS}}(\as) &=& - \left[ A(\as) + {\cal O}(\as^3)
\right] \ln N + {\cal O}(1) \;\;,\\
\gamma_{gg,N}^{\overline {\rm MS}}(\as) &=& - \frac{C_A}{C_F}
\;\left[ A(\as) + {\cal O}(\as^3)
\right] \ln N + {\cal O}(1) \;\;,
\eeeq
where $A(\as)$ is given in Eq.~(\ref{AB}) and
$\gamma_{ij,N}^{\overline {\rm MS}}(\as)={\cal O}(1/N)$ for $N \to \infty$ if
$i \neq j$.

Note, however, that the main feature of the Minimal Prescription, i.e. the
absence of factorially-growing coefficients, remains valid in any
factorization scheme. The only difference in using different schemes is in the
fact that one can introduce non-perturbative corrections which, although
suppressed by more than any power of $\Lambda/Q$ at fixed $1 - \tau$,
can actually have a different overall magnitude. As long
as $Q^2$ is sufficiently perturbative and $\tau$ sufficiently far from the
hadronic threshold, this difference should not sizeably affect the actual
value of the cross section. Obviously, approaching the essentially
non-perturbative regime $(Q^2 \to 1~{\rm GeV}^2, \tau \to 1)$, a physically
motivated treatment of non-perturbative effects has to be introduced. This is
beyond the scope of the present paper.
\section{Conclusions}
In our study of resummation procedures for partonic-threshold corrections,
we have found that the
factorial growth due to the infrared renormalon is only
a minor problem, when compared with the large spurious factorial
growth that is generated when the $x$ space
resummation formulae are limited to the LL or NLL level.
Although we consider the recent indication of the
cancellation of $1/Q$ renormalon effects in Drell--Yan pair production
\cite{BenekeBraun} as an important progress,
we find this spurious factorial growth
to be a much more serious problem in practice.
In fact, while a $1/Q$ power ambiguity in
the Tevatron top production cross section should be below the per cent level,
the ambiguity induced by the spurious factorial growth is at the level of 10\%.

In the present paper we have proposed the Minimal Prescription for the
resummation of partonic-threshold effects in hadronic processes.
This formula has a perturbative expansion free of factorially growing terms.
The ambiguity arising from its asymptotic nature is in fact exponentially
suppressed, behaving as $\exp(- H(1-x)Q/\Lambda)$.
We would also like to remark
that certain kinematic constraints
are respected by the MP formula. For example, for $N=1$ the MP
formula gives no resummation corrections. In general, sum rules
associated to low moments receive small corrections in our procedure.

As far as our phenomenological results are concerned, we have found
that in heavy flavour production, in current experimental configurations,
resummation effects are negligible.
The process of  high mass dijet production at the Tevatron, being much
closer to threshold, has instead non-negligible corrections.
We wish to remind the reader that, in this last case, in order
to perform a reliable phenomenological prediction,
resummation formulae should be applied not only to the production
process in question, but also to the processes that have been
used to extract the parton densities and to their evolution equations.
\vskip 0.5cm
\noindent {\bf Acknowledgements}
\par\noindent
We wish to thank Harris Contopanagos for useful discussions.
\appendix
\section{Asymptotic nature of the resummation formula}
\newcommand\gl{{<>}}
In this appendix we will study the asymptotic nature of our resummation
formula.
We are dealing with the LL function
\beq
G(\as)=\frac{1}{2\pi i}
\int^{C+i\infty}_{C-i\infty} dN\;\tau^{-N}\;h_N(\lambda)\,,\quad
h_N(\lambda)=\exp[\log N\,f(\lambda)]\,,\quad \lambda=\as\bzero\,\log N\,.
\eeq
Observe that we will consider $h_N(\lambda)$ as an independent function
of $N$ and $\lambda$.
We will focus upon the \MSB\ case, in which
\beq
f(\lambda)=\frac{c}{\lambda}\left[(1-2\lambda)\log(1-2\lambda)+2\lambda\right]
\eeq
with $c=A^{(1)}/(\pi\bzero)$,
but the proof can be easily generalized, since it only relies upon
the analyticity properties of $f(\lambda)$, and the fact that it does not grow
too strongly at infinity. Because of the same reason, the proof can also be
generalised to subleading logarithmic accuracy.

We introduce the notation
\beqn
h_N(\lambda)|_n&=&\sum_{i=0}^n h_N^{(i)}(0)\frac{\lambda^i}{i!}.
\nonumber \\
G(\as)|_n&=&\frac{1}{2\pi i}\int dN\;\tau^{-N}\;h_N(\lambda)|_n\;.
\eeqn
It is clear that $G(\as)|_n$ is the truncated expansion of $G(\as)$
to order $n$ in $\as$.
We will show that $G(\as)|_n$ is asymptotic to $G(\as)$, that is to say
that
\beq
\delta G(\as)=G(\as)-G(\as)|_{n-1}={\cal O}(\as^n).
\eeq
We begin by stating a few useful results for analytic functions.
For a generic analytic function $F(z)$ we have the identity
\beqn &&
 F(z)-\sum_{k=0}^{n-1} F^{(k)}(0)\,\frac{z^k}{k!}\;
=\int_0^z dx_1 \int_0^{x_1} dx_2 \ldots \int_0^{x_{n-1}} dx_n  F^{(n)}(x_n)
\nonumber \\\label{Taylorrest}
&& =\frac{1}{(n-1)!}\int_0^z (z-x)^{n-1}  F^{(n)}(x)\; dx\,.
\eeqn
where we assume that the integrations are performed along a straight
path from $0$ to $z$.
We then use Cauchy's inequality, which states that for a function
$F(z)$, analytic in a circle of radius $r$ around the point $z_0$
we have
\beq \label{Cauchy}
|F^{(n)}(z_0)|\leq\frac{n!}{r^n}\max_{|z-z_0|<r} |F(z)|\,.
\eeq
Assuming that $F(z)$ is analytic in a circle of radius $R$ around $z=0$,
from eq.~(\ref{Cauchy}) it follows that
\beq
|F^{(n)}(z)|\leq\frac{n!}{(R-|z|)^n}\max_{|w|<R} |F(w)|\,,
\eeq
which together with eq.~(\ref{Taylorrest}) yields the following bound
\beq
\left| F(z)-\sum_{k=0}^{n-1} F^{(k)}(0)\,\frac{z^k}{k!}\right|\;
\leq \; n\;\max_{|w|<R} |F(w)|\;
\int_0^{|z|} \frac{(|z|-x)^{n-1}}{(R-x)^n}\; dx\;,
\label{Taylorbound}
\eeq
which holds for $|z|<R$.
We now use the relation
\beq
\int_0^t \frac{(t-y)^{n-1}}{(R-y)^n}\;dy =
\left(\frac{t}{R}\right)^{n}\int_0^1 \frac{(1-y)^{n-1}}{(1-yt/R)^n}\;dy
 < \frac{1}{n}\left(\frac{t}{R}\right)^{n}\frac{R}{R-t}\;,
\eeq
(where the last inequality becomes an identity in the limit $n\to \infty$)
and get the bound
\beq
\left| F(z)-\sum_{k=0}^{n-1} F^{(k)}(0)\,\frac{z^k}{k!}\right|\;
\leq \; \max_{|w|<R} |F(w)|\;\frac{R}{R-z}\;
\left(\frac{z}{R}\right)^{n}\;,
\label{Ourbound}
\eeq
which we will apply in the following.

Consider now our expression for $\delta G$
\beq
\delta G(\as)=\frac{1}{2 \pi i}\int_{C-i\infty}^{C+i\infty}
\;dN\;\tau^{-N}\;
(h_N(\lambda)-h_N(\lambda)|_{n-1}) \; .
\eeq
We will apply the bound~(\ref{Ourbound}) to $h_N(\lambda)$, considered
as a function of $\lambda$ at fixed $N$. This function is analytic
up to the Landau pole at $\lambda=1/2$, so that in our case $R=1/2$.
We bent the $N$ contour towards the negative real axis,
as depicted in fig.~\ref{contour1}, with $N_0$ of order 1.
Choosing any real $0<r<1$ (close to $1$) we split our integral into a
near ($|\log N|<r/(2\as\bzero)$) and far ($|\log N|>r/(2\as\bzero)$) part,
\beqn
G(\as) &=& G_<(\as)+G_>(\as)
\nonumber\\
G_{\gl}(\as)&=&\frac{1}{2\pi i}
\int_{|\log N|\gl r/(2\as\bzero)} dN\;\tau^{-N}\;h_N(\lambda)\,.
\eeqn
For the near part, using eq.~(\ref{Ourbound}), we obtain immediately
\beqn &&
|\delta G_<|=\left| \int_{|\log N|<r/(2\as\bzero)}
dN\;\tau^{-N}\;(h_N(\lambda)-h_N(\lambda)|_{n-1}) \right|\leq
\nonumber \\&&
\int_{|N|\leq e^{r/(2\as\bzero)}} |dN\;\tau^{-N}|\; \frac{\exp[C|\log N|]}{1-r}
  \left(2\as\bzero\,|\log N|\right)^n\leq g_n \as^n\,,
\eeqn
where
\beq
g_n=\int |dN\;\tau^{-N}|\; \frac{\exp[C|\log N|]}{1-r}
  \left(2\bzero\,|\log N|\right)^n
\eeq
and
\beq
C=\max_{|\lambda|<r/2}|f(\lambda)|.
\eeq
By inspection it is easy to see that the far part is exponentially
suppressed. In fact, we can write
\beqn
|\delta G_>|&=&\left| \int_{|\log N|>r/(2\as\bzero)}
dN\;\tau^{-N}\;(h_N(\lambda)-h_N(\lambda)|_{n-1}) \right|
\nonumber \\ &\leq&
\int_{|\log N|>r/(2\as\bzero)} |dN|e^{-\log\frac{1}{\tau}|N|}
\;\left| h_N(\lambda)-h_N(\lambda)|_{n-1}\right| \;.
\eeqn
Since $f(\lambda)$ is logarithmically bounded in the far region,
$h_N(\lambda)$ grows at most as $(N)^{\ln \ln N}$.
Instead, $h_N(\lambda)|_n$ is polynomial in $\log N$. Therefore,
the dominant factor in the integrand
is the exponential, which is of the order of
$\exp(-\exp(r/(2\as\bzero))\log 1/\tau)$.
We have therefore proved that the MP formula is asymptotic to its formal $\as$
expansion.

It is easy to find, using saddle point
methods, the rate of growth of $g_n$. We find, for $n\ll\log 1/\tau$,
the following power behaviour
\beq
g_n\approx \left(2\bzero\,\log\frac{C}{\log 1/\tau}\right)^n\,,
\eeq
and for $n\gg\log 1/\tau$ we obtain the coefficients
\beq
g_n\approx \left(2\bzero\,\log\frac{n}{\log 1/\tau}\right)^n\,,
\eeq
whose growth is faster than any power, but much slower than factorial.
We can estimate the value of $n$ at which $\as^n g_n$ has a
minimum from the equation
\beq\label{minimumcond}
\frac{\as^{n+1} g_{n+1}}{\as^n g_n}\approx
\left(\frac{\log\frac{n+1}{\log 1/\tau}}{\log\frac{n}{\log 1/\tau}}
\right)^n 2\as\bzero\;\log\frac{n+1}{\log 1/\tau}\approx
e^{\frac{1}{\log\frac{n}{\log 1/\tau}}}\;
2\as\bzero\;\log\frac{n+1}{\log 1/\tau}=1\;.
\eeq
The value at the minimum is obtained from
\beq
\as^n g_n \approx  \exp\left(n\log\left[2\as\bzero\,
\log\frac{n}{\log 1/\tau}\right]\right)
\eeq
by using eq.~(\ref{minimumcond}). We get
\beq\label{minimumval0}
\as^n g_n \approx \exp\left(-\frac{n}{\log\frac{n}{\log 1/\tau}}\right)\,.
\eeq
The value of $n$ at the minimum is
\beq
n \approx e^\frac{1}{2\as\bzero}\;\log\frac{1}{\tau}\,,
\eeq
which inserted in eq.~(\ref{minimumval0}) yields
\beq
\as^n g_n \approx \exp\left(-2\as\bzero\,e^\frac{1}{2\as\bzero}\,
\log\frac{1}{\tau}\right)\;.
\eeq
Remembering that $\alpha_S\,\bzero=1/\log(Q^2/\Lambda^2)$ we see that,
taking $r\to 1$,
these suppression factors correspond to exponentials of the form
(neglecting logarithmic factors in the exponent)
\beq
e^{-H\frac{Q(1-\tau)}{\Lambda}} \; ,
\eeq
which is suppressed by more than any power of $Q$.
Notice that as $Q(1-\tau)$ approaches $\Lambda$ the above expression
becomes of order 1. In fact, in this limit, the total mass of the radiation
that accompanies the production of the object with mass $Q$ becomes of
order $\Lambda$. This is clearly a regime over which we do not have
any perturbative control.

As a last point, we discuss the inclusion of other $N$-dependent
factors in the integrand like the partonic cross section
in the case of heavy flavour production, or the parton luminosities.
Partonic cross sections have a negative-power behaviour in $N$ for large $N$,
so they will appear in the integrand as factors of $\exp[-M\log N]$,
which simply modifies the value of the constant $C$. Similar effects
would be given by common parametrizations of structure functions.
Of course, we cannot guarantee that the structure functions themselves
do not have a Sudakov like behaviour at large $x$.
The only statement we can make is that in the \MSB\ scheme,
if such a behaviour is not present in the initial conditions,
it will not arise because of evolution (see sec.~\ref{remark}).

\section{Numerical implementation of the MP formula}

In principle, there are no difficulties in the numerical implementation of the
MP formula. However, the modern sets of parton densities are usually provided
in terms of numerical codes or of parametrized expressions which, in practice,
cannot be used to evaluate their $N$-moments in analytic form for arbitrary
complex values of $N$. To overcome this practical difficulty we have to
rewrite the MP formula in terms of an $x$-space convolution of the parton
densities and the inverse Mellin transformation of the coefficient factor
$\Delta_N$.
Since $\Delta_N(Q^2)$ in formula~(\ref{MP})
has singularities to the right of the integration contour, much care
has to be taken when turning formula ~(\ref{MP}) to an $x$-space integral.
We begin by observing that, for large $N$, $\Delta_N(Q^2)$ (in the
$\MSB$ scheme) is suppressed by more
than any power of $N$ as $N\to\infty$.
We will also assume that $F_N$ is suppressed by some powers of $N$ as
$N\to\infty$. We will limit the present discussion to the Drell--Yan
case, although the extension to other cases is straightforward.

We rewrite formula~(\ref{MP}) by setting
\beq
F^2_N(Q^2)=\int_0^1\frac{dx}{x}\,x^N\,{\cal L}(x,Q^2)
\eeq
where
\beq
{\cal L}(x,Q^2)=\int_x^1\,F(x/z)\,F(z)\,\frac{dz}{z}\,.
\eeq
We obtain
\beqn
\sigma(\tau)&=&\frac{1}{2\pi i}
\int_{C_{\rm MP}-i\infty}^{C_{\rm MP}+i\infty}\; \tau^{-N}\;
\Delta_N(Q^2)\; \left[\int_0^1\,\frac{dx}{x}\,x^N{\cal L}(x,Q^2)\right]
\,dN\;
\nonumber \\
&=& \int_0^1\frac{dx}{x}\,{\cal L}(x,Q^2)\;\frac{1}{2\pi i}
\int_{C_{\rm MP}-i\infty}^{C_{\rm MP}+i\infty}\;
\left(\frac{\tau}{x}\right)^{-N}
\;\Delta_N(Q^2)\,dN\;.
\label{xMPform}
\eeqn
Observe that the $x$ integration extends from 0 to 1, not from $\tau$
to 1. This is because of the fact that, due to the Landau singularity
to the right of the contour, the Mellin transform of $\Delta_N(Q^2)$
does not vanish when its argument is greater than 1.
It does however vanish very fast when its argument is above 1 by an
amount greater than $\Lambda/Q$, so that the basic parton model
assumptions are not violated.
Let us in fact call $z=\tau/x$. For $z>1$ we have
\beq
\frac{1}{2\pi i}
\int_{C_{\rm MP}-i\infty}^{C_{\rm MP}+i\infty}\;\Delta_N(Q^2)\;z^{-N}\; dN\;
=\frac{1}{\pi}\int_{N_L}^\infty e^{-N\log z}\;\Im \Delta_{N+i\ep}(Q^2)\;dN\;.
\eeq
Using the formula
\beq
e^{-N t}=\sum_{k=0}^\infty\;\frac{(-1)^k}{N^{k+1}}\;\delta^{(k)}(t)\;,
\eeq
where $\delta^{(k)}(t)$ is the $k^{\rm th}$ derivative of the $\delta$
function, we obtain
\beqn
&&\frac{1}{2\pi i}
\int_{C_{\rm MP}-i\infty}^{C_{\rm MP}+i\infty}\;\Delta_N(Q^2)\;z^{-N}\; dN\;
=\;\delta(\log z)\frac{1}{ \pi  }
\int_{N_L}^{\infty}\;\Im \Delta_{N+i\ep}(Q^2)\; \frac{dN}{N}\;
\nonumber \\ \label{DeltoidRemainder}
&&-\delta^{(1)}(\log z)\frac{1}{\pi}
\int_{N_L}^{\infty}\;\Im\Delta_{N+i\ep}(Q^2)\;
\frac{dN}{N^2} +\ldots\;.
\eeqn
Using saddle point arguments it is possible to show that
the first term is of order 1, while the subsequent terms are suppressed
by powers of $\Lambda/Q$. Let us in fact consider the integral
\beq
I_m=\frac{1}{\pi}\int_{N_L}^{\infty}\;\Im\Delta_{N+i\ep}(Q^2)\;
\frac{dN}{N^{m+1}}=\frac{1}{2\pi i}
\int_{C_{\rm MP}-i\infty}^{C_{\rm MP}+i\infty}
\Delta_{N}(Q^2)\;\frac{dN}{N^{m+1}} \; .
\eeq
The $N$ integration can be turned into a $\lambda=\as\bzero\log N$ integration,
where the contour ${\cal C}$ comes from $\infty - \ep i$, encircles clockwise
the Landau singularity at $\lambda=1/2$, and then goes to  $\infty + \ep i$.
In the \MSB\ scheme we have
\beq
I_m=
\frac{1}{\bzero\as}\frac{1}{2\pi i}\int_{\cal C}
\exp\left(\frac{r}{\as \bzero}\left[2\lambda-\frac{m}{r} \lambda
+(1-2\lambda)\log(1-2\lambda)\right]\right)
\;d\lambda\;,
\eeq
where $r=A^{(1)}/(\bzero\pi)$.
The integration contour can be deformed into a steepest descent
contour, where the saddle point is located at
$\lambda=\frac{1}{2}(1-e^{-\frac{m}{2r}})$, and the saddle point
estimate of the integral is
\beqn
I_m &\approx&
\exp\left(\frac{r}{\as \bzero}\left[1-\frac{m}{2r}-e^{-\frac{m}{2r}}\right]
\right)\frac{1}{\bzero\as}\frac{1}{2\pi i}\int_{-i\infty}^{i\infty}
\exp\left(\frac{2 r}{\as \bzero}e^{\frac{m}{2r}}\;\bar\lambda^2\right)
\;d\bar\lambda
\nonumber \\
&=& e^{-\frac{m}{4r}}\sqrt{\frac{1}{8\pi r\as \bzero}}
\exp\left(\frac{r}{\as \bzero}
\left[1-\frac{m}{2r}-e^{-\frac{m}{2r}}\right]\right)\;.
\eeqn
We have verified numerically that the above estimate
is in fact quite good, although terms down by a power of $\as$ have
been neglected in the saddle point approximation.
It is apparent from the formula that, except for $m=0$, $I_m$ is
power-suppressed, with an exponent increasing with $m$.
In our philosophy, however, all the terms of the expansion should be kept,
since we have chosen not to include any power-suppressed effects
in our definition.

In principle there is nothing wrong in the integral (\ref{xMPform}),
it is convergent, and it should be possible to perform it
numerically. In practice, however, this is not convenient.
Even if a careful importance sampling is implemented in order
to correctly treat the region around $z=1$, the integrand performs
a few oscillations, in which large cancellations occur, so that
a direct numerical integration proves in this case to be too
time-consuming.
In order to overcome this problem we use the following
trick. We introduce a fake luminosity ${\cal L}^{(0)}(x,\tau)$,
defined as
\beq
{\cal L}^{(0)}(x,\tau)=A\,(1-x)^\alpha\, x^{-\beta}\;\left(1-P(x)\right)\;.
\eeq
where $P(x)$ is a polynomial in $x$. We then define
\beq
x_i\equiv\tau+(i-1)\eta\;,\quad l_i\equiv{\cal L}^{(0)}(x_i,Q^2)\;,
\eeq
where $i=1,\ldots,4$ and $\eta$ is a small quantity, usually taken to be
of the order of $\Lambda/Q$.
We then choose
\beq
P(x)=B\,(x-x_1)\,(x-x_2)\,(x-x_3)\;,
\eeq
and $A$, $B$, $\alpha$, $\beta$ in such a way that
\beq
{\cal L}^{(0)}(x_i,\tau)=l_i\;,\quad i=1,\ldots,4\;.
\eeq
These conditions can be easily solved to yield
\beqn
\alpha&=&\dispfrac{\log\frac{l_1}{l_3}\log\frac{x_1}{x_2}
-\log\frac{l_1}{l_2}\log\frac{x_1}{x_3}}{
\log\frac{x_1}{x_2}\log\frac{1-x_1}{1-x_3}
-\log\frac{1-x_1}{1-x_2}\log\frac{x_1}{x_3}}
\nonumber \\
\beta&=&\dispfrac{
\log\frac{l_1}{l_3}\log\frac{1-x_1}{1-x_2}
-\log\frac{l_1}{l_2}\log\frac{1-x_1}{1-x_3}}
{
\log\frac{x_1}{x_2}\log\frac{1-x_1}{1-x_3}
-\log\frac{1-x_1}{1-x_2}\log\frac{x_1}{x_3}}
\nonumber \\
A&=&\frac{l_1}{(1-x_1)^\alpha\,x_1^{-\beta}}
\nonumber \\
B&=&\frac{A\,(1-x_4)^\alpha\,x_4^{-\beta}\;-l_4}
{A\,(1-x_4)^\alpha\,x_4^{-\beta}
\;(x_4-x_1)\,(x_4-x_2)\,(x_4-x_3)}\;.
\eeqn
We now rewrite formula (\ref{xMPform}) in the following way
\beqn
\sigma(\tau)&=& \int_0^1\frac{dx}{x}\,\left({\cal L}(x,Q^2)
-{\cal L}^{(0)}(x,\tau)\right)\;
\frac{1}{2\pi i}
\int_{C_{\rm MP}-i\infty}^{C_{\rm MP}+i\infty}\;
\left(\frac{\tau}{x}\right)^{-N}
\;\Delta_N(Q^2)\,dN
\nonumber \\
&&+\frac{1}{2\pi i}
\int_{C_{\rm MP}-i\infty}^{C_{\rm MP}+i\infty}\;
\tau^{-N}\;{\cal L}^{(0)}_N(\tau)
\;\Delta_N(Q^2)\,dN\;,
\label{xMPformmod}
\eeqn
where as usual
\beq
{\cal L}^{(0)}_N(\tau)=\int_0^1 \,\frac{dx}{x}\;x^N\;{\cal L}^{(0)}(x,\tau)\;,
\eeq
which is easily performed analytically for possibly complex values of $N$.
The difficult region of $x\to\tau$ is now strongly suppressed
in the first integral of eq.~(\ref{xMPformmod}), and the whole formula
can be implemented numerically with no further difficulties.
The only problem one encounters is instabilities with structure
function packages that perform piecewise interpolation of tables,
since they do not give a smooth structure function. Thus, we found it
much easier to use the CTEQ sets, which are analytic in $x$,
rather than the MRS sets, which use an interpolation procedure.

\section{Some formulae and theorems on the Mellin transforms}
Let us begin by discussing the basic result
\beq\label{ntox}
\int_{C-i\infty}^{C+i\infty}\frac{dN}{N}\,
e^{-N\log x + \log N \,F(\as\log N)}=
\theta(1-x)\;e^{ l  \,F(\as l ) }
\times [1 + E(\as,l)]
\eeq
where
\beqn
l&=&-\log(-\log x)\approx \log\frac{1}{1-x}\;,
\nonumber \\
F(z)&=&\sum_{k=1}^\infty f_k\;z^k
\nonumber \\
E(\as,l)&=& \sum_{k=1}^\infty \as^k \sum_{j=0}^k e_{kj}\;l^j\;.
\eeqn
Therefore, the $E$ term is a NLL correction, since its power expansion
has no terms with more powers of $l$ than of $\as$.
Equation~(\ref{ntox}) should be interpreted in the sense of its
formal power expansion in $\as$. A simple proof has been given in
ref.~\cite{CataniWebber}. For completeness, we shortly report here
the basic argument. We rewrite the integral in terms of $z=-N\log x$
\beqn
\int\frac{dN}{N}\,
e^{-N\log x + \log N \,F(\as\log N)}&=&
\int\frac{dz}{z}\,
e^{z + [\log z + l] \,F(\as[\log z + l])}
\nonumber\\
&=&e^{ l \,F(\as l) } \times
\int\frac{dz}{z}\,
e^{z + G(\log z,\as,l)}\;,
\eeqn
where
\beq
G(\log z,\as,l)=[\log z + l] \,F(\as[\log z + l])-l \,F(\as l)\;
= \sum_{k=1}^\infty \as^k \sum_{j=0}^k g_{kj}(\log z) \;l^j\;,
\eeq
where the coefficients $g_{kj}(\log z)$ are polynomials in $\log z$.
Therefore $G$ represents a NLL correction.
Expanding its exponential, and integrating term by term in $z$,
we get precisely an expression of the form $1+E(\as,l)$, which is still
a NLL correction.
By taking the derivative of both sides of eq.~(\ref{ntox})
with respect to $\log x$ we get
\beq\label{ntox1}
\frac{1}{2\pi i}\int_{C-i\infty}^{C+i\infty} dN\,
e^{-N\log x + \log N \,F(\as\log N)}=
x\frac{d}{dx} \left\{ \theta(1-x)\;e^{l  \,F(\as l ) }
\times [1 + E(\as,l)] \right\} \; ,
\eeq
which has the form used in section~3.

In section~\ref{Problems}, we encounter the Mellin transforms
\beq
\frac{1}{2\pi i}\int d N\;x^{-N}\;\log^m N\;.
\eeq
We have
\beq
R_{m,k}(x)\equiv \int\frac{dN}{2\pi i}\,x^{-N}\frac{\log^m N}{N^k}=
(-)^m\frac{d^m}{d k^m}\;T(k,x),
\eeq
where
\beq
T(k,x)=\int \frac{dN}{2\pi i}x^{-N}e^{-k\log N}=\frac{1}{\Gamma(k)}
\left(\log\frac{1}{x}\right)^{k-1}\,,\quad {\rm Re}\,k>0\,.
\eeq
To prove the above formula, observe that it can always be interpreted
according to a contour distortion, as the one given in fig.~\ref{contour1}.
If $k <1$, the contour can be made to coincide with the negative real axis,
without convergence problems:
\beqn
I(k)&=&-\int\; dt\; x^t \; \frac{e^{-k(i\pi+\log t)}-\mbox{cc}}{2\pi i}
\nonumber \\
&=&\frac{\sin(k\pi)}{\pi}\int dt e^{-t\log\frac{1}{x}} t^{-k}
\nonumber \\
&=&\left(\log\frac{1}{x}\right)^{k-1}
\frac{\sin(k\pi)\,\Gamma(1-k)}{\pi}
\nonumber \\
&=&\frac{1}{\Gamma(k)}\left(\log\frac{1}{x}\right)^{k-1}
\eeqn
(using $\Gamma(1-x)\Gamma(x)\sin(\pi x)=\pi$).
This proof is valid for $k<1$. On the other hand, $I(k)$ can also
be given in a form that is manifestly analytical for all ${\rm Re}\,k >0$,
by distorting the contour away from $N=0$. By analytic continuation we
can therefore extend the above result to all the ${\rm Re}\,k >0$ region.
We finally get the result
\beq
R_{m,k}(x)=
\left(-\frac{\partial}{\partial k}\right)^m\; e^{(k-1)\log\log\frac{1}{x}
-\log\Gamma (k)}\;.
\eeq
From the above formula we also obtain the property
\beq
\int_0^1\,dx\,R_{m,k}(x)=0\;,
\eeq
valid for $k,m\, >\, 0$, which follows from
\beq
\int_0^1\log^{k-1}\left(\frac{1}{x}\right)\;dx\;=\;\Gamma(k)\,.
\eeq
Notice now that
\beq
R_{m,1}(x)=\left(-\log\log\frac{1}{x}\right)^m -
m \left(-\log\log\frac{1}{x}\right)^{m-1} \gamma_E
+\ldots\;.
\eeq
If we keep only the leading logarithmic term in the expansion,
we get
\beq
\int_0^1 R_{m,1}(x) \;dx\approx \int_0^1\left(-\log\log\frac{1}{x}\right)^m\;
 dx
\approx m!\;.
\eeq
Observe however that also the integral of the first subleading term
is proportional to $m!$, with opposite sign. In fact, if we keep
all subleading terms we must get zero, as shown earlier.
These are precisely the dangerous, factorially growing terms that
arise when neglecting next-to-leading terms in the $x$-space resummation
formulae.

\end{document}